\documentclass{jpp}
\usepackage{graphicx}

\usepackage[utf8]{inputenc}
\usepackage[T1]{fontenc}
\usepackage{amsmath}

\usepackage[bbgreekl]{mathbbol}
\usepackage{amssymb}
\usepackage{amsfonts}
\usepackage{bm}
\usepackage{bbm}
\usepackage{subcaption} % Required for subfigure environment
\usepackage{upgreek}
\usepackage{url}
\usepackage{comment}

\newtheorem{remark}{Remark}

\shorttitle{Geometric numerical discretization of electromagnetic quasineutral models}
\shortauthor{N. Narechania, E. Poulsen and E. Sonnendr\"ucker}

\title{Geometric numerical discretization of electromagnetic quasineutral models}

\author{Nishant Narechania\aff{1}
  \corresp{\email{nishant.narechania@ipp.mpg.de}},
  Emil Poulsen\aff{1}
 \and Eric Sonnendr\"ucker\aff{1,2}}

\affiliation{\aff{1}Max Planck Institute for Plasma Physics, Garching, Germany
\aff{2}School of Computation Information and Technology, Technical University of Munich, Garching, Germany}

\newcommand{\fracp}[2]{\frac{\partial #1}{\partial #2}}
\newcommand{\fract}[2]{\frac{\mathrm{d} #1}{\mathrm{d} #2}}
\newcommand{\dd}{\mathop{}\!\mathrm{d}} %for integrals

\newcommand{\ii}{\mathop{}\!\mathrm{i}}

\newcommand{\half}{\frac{1}{2}}
\newcommand\bk{\boldsymbol{k}}
\newcommand\bX{\boldsymbol{X}}

\newcommand\bV{\boldsymbol{V}}

\newcommand\bx{\boldsymbol{x}}
\newcommand\bv{\boldsymbol{v}}
\newcommand\bn{\boldsymbol{n}}
\newcommand\bE{\boldsymbol{E}}
\newcommand\bB{\boldsymbol{B}}
\newcommand\bA{\boldsymbol{A}}
\newcommand\bJ{\boldsymbol{J}}
\newcommand\bQ{\boldsymbol{Q}}

\newcommand\bD{\boldsymbol{D}}
\newcommand\bH{\boldsymbol{H}}
\newcommand\bK{\boldsymbol{K}}
\newcommand\bL{\boldsymbol{L}}
\newcommand\bM{\boldsymbol{M}}
\newcommand\bN{\boldsymbol{N}}

\newcommand\cR{\mathcal{R}}
\newcommand\cC{\mathcal{C}}

\newcommand{\arr}[1]{{\bm{\mathsf{#1}}}}

\newcommand\arrA{{\bm{\mathsf{A}}}}
\newcommand\arrB{{\bm{\mathsf{B}}}}

\newcommand\arrE{{\bm{\mathsf{E}}}}

\newcommand\arrJ{{\bm{\mathsf{J}}}}

\def\matd{\mathbb{d}}

\def\matC{\mathbb{C}}
\def\matD{\mathbb{D}}

\def\matG{\mathbb{G}}
\def\matH{\mathbb{H}}

\def\matI{\mathbb{I}}

\def\matO{\mathbb{O}}

\newcommand{\tA}{\ensuremath{\delta{\boldsymbol{A}}}}
\newcommand{\tphi}{\ensuremath{\delta{\phi}}}

\usepackage{xcolor}
\newcommand{\nmn}[1]{{\color{black} #1}}

\begin{document}

\maketitle

\begin{abstract}
 In this work, the geometric electromagnetic Particle-in-Cell (PIC) framework, \texttt{GEMPICX}, is extended to solve the quasineutral, fully kinetic Vlasov-Maxwell equations on dual grids using mimetic finite differences. The discrete action principle is derived, taking into account the duality between the grids. The temporal derivative of the electric field does not directly appear in the dynamical system for the quasineutral model. Hence, a discretized curl-curl equation is used to implicitly obtain the electric field at every time-step. This also circumvents the need to obtain electric potentials. A Lagrange multiplier is used to maintain the discretized divergence of the current density at machine zero.
\end{abstract}

%%%%%%%%%%%%%%%%%%%%%%%%%%%%%%%%%%%%%%%%%%%%%%%%%%%%%%%%%
\section{Introduction}\label{sec:intro}

The study of magnetized plasmas is of crucial importance in applications such as fusion energy, space weather and astrophysics. Numerical simulations provide an effective way of studying plasma behavior, by avoiding the prohibitive costs of conducting in situ and remote measurements. The large variation in length and time scales in the dynamics of magnetized plasmas poses a challenge for numerical methods and computer simulations (\citet{brambilla1998kinetic, fitzpatrick2022plasma, stix1992waves}). To capture the diverse range of phenomena observed in magnetized plasmas, researchers have proposed numerous physical models over the years, with various numerical techniques based on these models. These models are broadly divided into three categories: kinetic models that describe the full velocity-space distribution function of each species; fluid models assuming a near-Maxwellian distribution, and therefore allowing the plasma to be characterized by moments of the distribution function such as density, temperature and bulk velocity; and hybrid models that are a combination of the other two. Kinetic models are used when kinetic effects such as non-Maxwellian distributions, wave-particle interactions and kinetic instabilities become important. They employ the Vlasov equation in combination with Maxwell's laws of electromagnetism to describe the dynamics of the plasma. When the physical dimensions of the problem studied are much larger (> $\sim$20 times) than the Debye length, the plasma can be considered as being quasineutral across these length scales i.e. having an equal number of positive and negative charge carriers. Quasineutral models filter out the lower length scales over which charge separation phenomena occur. They also remove the description of high-frequency electromagnetic wave propagation from the Vlasov-Maxwell system. Quasineutral models are therefore reduced models that eliminate the need for resolving these small time and length scales, resulting in faster numerical schemes and more efficient use of computational resources.

%\citet{hewett1994low} proposed a finite-difference algorithm for a quasineutral hybrid-kinetic model. In this model, called a Darwin model, the electrons can have a zero or finite mass and are treated as a neutralizing background fluid whereas the ions are fully kinetic. \citet{Sonnendrucker1995} developed a finite-element implementation of the Darwin-PIC model for unstructured meshes.

Numerical algorithms for quasineutral systems have been developed over the years and still remain an active area of research in computational plasma physics. \citet{joyce1997electrostatic} developed an electrostatic PIC technique for quasineutral plasmas wherein the electric field was obtained directly from the momentum equation and the quasineutrality condition, rather than the Poisson equation. \citet{cheng1999} introduced a kinetic-fluid model wherein multiple ion species were modeled as a single fluid, and kinetic effects were incorporated through plasma pressure tensors that were computed from particle distribution functions obtained by solving the Vlasov or gyrokinetic equation. \citet{crispel2005quasi} developed numerical models for the quasineutral limit of a two-fluid, isentropic Euler-Poisson system. They later proposed for this system an improved, asymptotic-preserving (AP) scheme that was less restricted in terms of temporal and spatial discretization \citep{crispel2007asymptotic}. \citet{degond2006asymptotically} introduced the first asymptotically stable PIC scheme for the near-quasineutral Vlasov-Poisson system for electrostatic, collisionless plasmas. They later made this scheme fully asymptotic-preserving and extended its stability to larger timesteps \citep{degond2010asymptotic}. Degond \emph{et. al.} then also developed AP schemes for fluid-like models such as the Euler-Maxwell model \citep{degond2012numerical} and hybrid systems like the Euler-Poisson-Boltzmann equations \citep{degond2012numerical1}, in the quasineutral limit. Finally, they developed AP PIC methods for the quasineutral Vlasov-Maxwell model, allowing for fully electromagnetic plasma simulations with significant magnetic fields \citep{Degond2017Asymptotic-Pres}. For a detailed review and comparison of these asymptotic schemes, the reader is referred to the work by Degond and Deluzet \citep{Degond2017AP-review}.

There has also been a considerable amount of work done in the plasma physics community on methods that preserve certain geometric structures of the physical system of equations being solved. These structures could be divergences, gauge invariances or conservation laws for instance. For instance, energy-conserving PIC discretizations of the electrostatic Vlasov-Poisson \citep{lewis1970energy} and electromagnetic Vlasov-Maxwell \citep{lewis1972variational} systems were developed that used a leapfrog time-integrator. These early works on structure-preserving methods used finite-differences for discretizing fields. \citet{markidis2011energy} proposed a fully discrete energy-conserving semi-implicit PIC scheme based on the implicit midpoint discretization of Maxwell’s equations. \citet{lapenta2017exactly} then reduced the computational cost of this method by removing the need for non-linear iterative solution procedure, while still maintaining discrete energy conservation. \citet{chen2011energy} proposed an energy-conserving fully implicit discretization for the electrostatic Vlasov-Poisson system. \citet{chen2015multi} developed a nonlinearly implicit electromagnetic PIC algorithm based on the Vlasov-Darwin model that exactly conserves both total energy and charge through a consistent field-particle coupling. However, these energy-conserving implicit schemes are not derived from a variational or Hamiltonian principle, and therefore do not preserve the de Rham or Poisson structure of the Vlasov-Maxwell system. Methods that rely on the de Rham structure of the Maxwell equations were also being developed for PIC methods that used finite-element meshes \citep{bossavit1998computational, hiptmair2002finite}. These ensured an exact preservation of the discretized version of the Gauss laws. \citep{tronci2014hybrid} investigated the conservation properties of the Hamiltonian structures of some quasineutral hybrid Vlasov-MHD models. More recently, a geometric electromagnetic PIC (GEMPIC) method that used finite elements, was developed, based on the Vlasov-Maxwell system's intrinsic Hamiltonian structure \citep{kraus2017gempic, campos2022variational}. Techniques from finite element exterior calculus (FEEC) \citep{arnold2018finite} were employed, which conserved discretized versions of the Gauss laws, total energy, the Poisson structure and Casimir invariants.

Besides finite elements, this approach has now also been developed with a discretization based on structure-preserving mimetic finite differences \citep{kormann2024}. These mimetic discretizations employ operators that exactly imitate vector calculus identities on discrete spaces \citep{bochev2006principles}. To achieve this, unknowns are represented as point values, edge integrals, face integrals, and volume integrals, constituting a discrete de Rham complex. Recently, this approach has also been extended to the drift-kinetic (DK) model and the hybrid model with drift electrons and kinetic ions, by \citet{meng2025}. In this work, we aim to extend this method to the quasineutral limit of the Vlasov-Maxwell PIC system with both kinetic electrons and ions. We start with the Lagrangian from \citet{Tronci2015Neutral-Vlasov-} and then propose a discretized Lagrangian based on the geometric structure-preserving discretization related to mimetic finite differences. Importantly, our approach is derived directly in the quasineutral limit, rather than relying on asymptotic consistency of a full Vlasov-Maxwell solver upon the Debye length approaching zero, as in the work by \citet{Degond2017Asymptotic-Pres}. Because the formulation is derived directly in the quasineutral limit, the scheme ensures the absence of electromagnetic and Langmuir waves in the numerical solution. On the contrary, asymptotic-preserving approaches rely on numerical damping to get rid of these waves in the quasineutral limit as the Debye length tends to zero. Another important feature of our model is that the discretized field equations and equations of motion for particles are both derived from the discretized Lagrangian, ensuring structure preservation in the semi-discretized space. At the semi-discrete level, we show that this system preserves the sum of the magnetic and kinetic energies, and maintains the zero-divergence requirement of the current density for the quasineutral model. We then check the wave spectra of single- and double-species quasineutral plasmas to verify our numerical model, using quasi-1D test cases with periodic boundary conditions. A wave-damping simulation with periodic boundary conditions is also conducted to verify the theoretically predicted damping rate obtained from the dispersion relation.

The organization of this paper is as follows: Section \ref{sec:qnvmmodel} below describes the quasineutral, fully kinetic Vlasov-Maxwell model, including the governing equations, action principle, and the electric field equation. Section \ref{sec:fielddiscretization} describes the spaces and discretization used for describing the fields. Section \ref{sec:particlediscretization} describes the splines used for the coupling between particles and fields. Section \ref{sec:discreteqnvmmodel} details the space-discretization for the governing equations derived from the discrete action principle, the semi-discretized electric field equation and energy-conserving properties. %\nmn{Section \ref{sec:stabilityanalysis} derives the time-step criterion for stability of the numerical time-stepping algorithm, using a cold plasma model.}
Section \ref{sec:timestepping} details the time-stepping algorithm used to solve the discretized quasineutral system. The dispersion relation and various waves obtained are described in Section \ref{sec:dispersionrelation}. Numerical tests with periodic boundary conditions used to validate the framework and scaling studies are detailed in Section \ref{sec:tests}. Finally, Section \ref{sec:conclusions} summarizes our findings and provides general conclusions.
%%%%%%%%%%%%%%%%%%%%%%%%%%%%%%%%%%%%%%%%%%%%%%%%%%%%%%%%%

%%%%%%%%%%%%%%%%%%%%%%%%%%%%%%%%%%%%%%%%%%%%%%%%%%%%%%%%%
\section{The quasineutral Vlasov-Maxwell model}\label{sec:qnvmmodel}

\subsection{Governing equations}\label{sec:goveqns}

In a fully kinetic model of a plasma, a species $s$ is described using a phase-space distribution function $f_s$ that is a function of space ($\bx$), velocity ($\bv$) and time ($t$). This function $f_s$ evolves according to the Vlasov equation given by
\begin{equation}\label{eq:qnVlasov}
    \fracp{f_s}{t} + \bv \cdot\nabla_x f_s +  \frac{q_s}{m_s}{(\bE + \bv \times \bB)} \cdot\nabla_v f_s=0,
\end{equation}
where $q_s$ and $m_s$ are the particle charge and mass for the species $s$, and $\bE$ and $\bB$ are the electric and magnetic fields. ($\bX$) and ($\bV$) are characteristics of the Vlasov equation. In the quasineutral model, the electric and magnetic fields satisfy the following Maxwell's equations:
\begin{gather}
    \nabla \times \bB = \mu_0 \bJ, \label{eq:qnampere}\\
    \fracp{\bB}{t} + \nabla \times \bE = {\bf 0}, \label{eq:qnfaraday}\\
    \rho = 0, \label{eq:qnrho}\\
    \nabla \cdot \bB = 0. \label{eq:qngaussB}
\end{gather}
Here, $\mu_0$ is the magnetic permeability of vacuum. Equations \eqref{eq:qnVlasov}--\eqref{eq:qngaussB} together comprise the Vlasov-Maxwell system in the quasineutral limit. The charge density, $\rho$, and the current density, $\bJ$, are obtained by integrating the zeroth and first moments of the distribution function $f_s$ over the velocity space:
\begin{gather}
    \rho = \sum_{s} q_s \int f_s d\bv, \label{eq:rho}\\
    \bJ = \sum_{s} q_s \int f_s \bv d\bv. \label{eq:J}
\end{gather}
In the full Vlasov-Maxwell system, without taking the quasineutral limit, the Amp\`ere equation i.e. equation \eqref{eq:qnampere}, also contains the displacement current term given by $\frac{1}{c^2} \fracp{\bE}{t}$, which can also be written as $\mu_0 \epsilon_0 \fracp{\bE}{t}$. Here $c$ is the speed of light and $\epsilon_0$ is the permittivity of vacuum. In the full system, equation \eqref{eq:qnrho} becomes $\rho = \epsilon_0 \nabla\cdot \bE$. The quasineutral model is obtained by taking the formal limit $\epsilon_0 \to 0$, thus leading to the above equations. Integrating the Vlasov equation \eqref{eq:qnVlasov} over the velocity space and using the $\rho = 0$ condition from equation \eqref{eq:qnrho}, gives the divergence-free condition for the current density:
\begin{equation}\label{eq:qncontinuity}
    \nabla \cdot \bJ = 0.
\end{equation}
The magnetic field can be written in terms of the magnetic vector potential $\bA$ as
\begin{equation}
    \bB = \nabla \times \bA.
\end{equation}
The Amp\`ere's equation \eqref{eq:qnampere} then can be written as
\begin{equation} \label{eq:curlcurlA}
\nabla \times \nabla \times \bA = \nabla(\nabla \cdot \bA) - \nabla^2\bA = \mu_0 \bJ.
\end{equation}
The quasineutrality condition, $\nabla \cdot \bJ = 0$, makes it convenient to use the Coulomb gauge condition
\begin{equation} \label{eq:gauge}
    \nabla \cdot \bA = 0.
\end{equation}
This further simplifies equation \eqref{eq:curlcurlA} to the Laplace equation
\begin{equation} \label{eq:laplaceA}
- \nabla^2\bA = \mu_0 \bJ.
\end{equation}
We note that the use of periodic boundary conditions in the test cases considered here allows a straightforward imposition of the Coulomb gauge condition in equation \eqref{eq:gauge}. This in turn allows the curl-curl operator to be reduced to a Laplace operator that is invertible up to an additional constant. Non-periodic boundary conditions would introduce the need for additional constraints on $\bA$ at domain boundaries that are consistent with the Coulomb gauge.

\subsection{Action principle}\label{sec:action}

In order to obtain the above governing equations in the continuous form, we start from the action principle developed by \citet{Tronci2015Neutral-Vlasov-}, which reads

\begin{multline}\label{eq:action}
	L(\bA(t),\phi(t), \bX_s(t),\dot{\bX_s}(t),\bV_s(t)) 
	= \sum_s\int \left((m_s \bV_s(t) + q_s\bA(t,\bX_s(t))) \cdot \dot{\bX_s}(t) \right. \\ \left. - \frac 12 m_s V_s^2- q_s \phi(t,\bX_s(t))\right) f_{s,0} \dd \bx_0  \dd \bv_0 
	 - \int \frac{1}{2\mu_0} \left|  \nabla\times\bA(t,\bx)\right|^2  \dd \bx, 
\end{multline}
where $\bE = - \fracp{\bA}{t} - \nabla\phi$ and $\bB = \nabla\times\bA$. The sum is over all particle species.

\begin{remark}
	In principle the constraint $\nabla\cdot\bJ=0$ follows from $\nabla\times\bB = \bJ$, but we will not be able to enforce this exactly numerically. For this reason, we need to enforce it with a Lagrange multiplier which will yield a slight modification of the electric field used for pushing the particles, so that the divergence of the current generated by the particles stays 0.
\end{remark}

The physical system can be derived from this action. The variations with respect to $\bX_s$ and $\bV_s$ yield the Euler-Lagrange equations for the particles, $\frac{\partial}{\partial t} \frac{\partial L}{\partial\dot{\bX}} = \frac{\partial L}{\partial \bX}$ and $\frac{\partial L}{\partial \bV} = \bf 0$. In this case, they become 
\begin{gather}
	\fract{\bX_s}{t} =\bV_s, \label{eq:Xdots}\\
	\fract{\bV_s}{t} = \frac {q_s}{m_s}\left( \bE + \bV_s\times\bB\right).\label{eq:Vdots}
\end{gather}
The solutions of equations \eqref{eq:Xdots} and \eqref{eq:Vdots} are the characteristic equations of the Vlasov equation. Taking the total derivative of $f_s$, and using equations \eqref{eq:Xdots}--\eqref{eq:Vdots}, gives the Vlasov equation for $f_s$.

On the other hand the variations in $\bA$ yield
\begin{equation}
    \int \bB \cdot \nabla\times\tA \dd \bx = \sum_s q_s \int \tA \cdot \bv f \dd\bx \dd\bv, \quad\forall\tA,
\end{equation}
which becomes in strong form
\begin{equation}\label{eq:AmpereQN}
	\nabla\times\bB = \mu_0\bJ = \mu_0 \sum_s\bJ_s = \mu_0 \sum_s q_s \int f_s \bv  \dd \bv.
\end{equation}
The variations in $\phi$ yield
\begin{equation}
 \sum_s q_s \int \tphi f_s \dd\bx \dd\bv = 0, \quad\forall\tphi
\end{equation}
which is the quasineutrality condition
\begin{equation}\label{eq:QN-condition}
	\rho = \sum_s q_s \int f_s \dd\bv =  0.
\end{equation}
Equation \eqref{eq:QN-condition} above also implies $\nabla \cdot \bJ = 0$ because of the continuity equation \eqref{eq:qncontinuity}.
Moreover, the Faraday's equation \eqref{eq:qnfaraday} follows directly from the definition of the fields from the potentials. Similarly, the Gauss law for magnetism \eqref{eq:qngaussB} follows from the definition of the magnetic field.

\subsection{The electric field equation}\label{sec:efieldeqn}

In the quasineutral model, without the displacement current $\frac{1}{c^2} \fracp{\bE}{t}$ in the Amp\`ere's equation, the electric field cannot be evolved using its time derivative. An equation is therefore required to obtain the electric field at any time instant. Taking the curl of \eqref{eq:qnfaraday} and the time derivative of \eqref{eq:qnampere} and combining the two equations, we obtain
\begin{equation} \label{eq:qncurlcurl}
    \nabla \times \nabla \times \bE = - \mu_0 \fracp{\bJ}{t}.
\end{equation}
Next, multiplying the Vlasov equation by $q_s\bv$ and integrating over the velocity space, we obtain
\begin{equation}\label{eq:qndjdt}
    \fracp{\bJ}{t} = \sum_s\fracp{\bJ_s}{t} = \sum_s \frac{q_s}{m_s} (\rho_s\bE + \bJ_s \times \bB - \nabla \cdot \mathbb{S}_s).
\end{equation}
where $\rho_s$, $\bJ_s$ and $\mathbb{S}_s$ are the contributions of species $s$ to the charge density, current density and stress tensor, given by

\begin{gather}
    \rho_s =  q_s \int f_s d\bv, \label{eq:rho_s}\\
    \bJ_s = q_s \int f_s \bv d\bv, \label{eq:J_s}\\
\mathbb{S}_s = m_s \int f_s \bv \otimes \bv d\bv. \label{eq:S_s}
\end{gather}
Substituting equation \eqref{eq:qndjdt} in equation \eqref{eq:qncurlcurl} gives the following equation for the electric field
\begin{equation}\label{eq:qnEeqn}
    \nabla \times \nabla \times \bE + \mu_0 \sum_s \frac{q_s}{m_s}\rho_s\bE = \mu_0\sum_s \frac{q_s}{m_s} (-\bJ_s \times \bB + \nabla \cdot \mathbb{S}_s).
\end{equation}
This equation can be solved at every time-step to obtain the electric field. It must be noted that we also have the option to solve equation \eqref{eq:qncurlcurl} to obtain $\bE$. However, in the discrete setting, this would require the use of a time discretization scheme to obtain the discrete $\fracp{\bJ}{t}$. The electric field solution would depend on the discrete derivative used, which is not the case for equation \eqref{eq:qnEeqn}. More importantly, the system in equation \eqref{eq:qncurlcurl} is singular due to the curl-curl term. On the other hand, equation \eqref{eq:qnEeqn} is non-singular and determinate on account of the $\sum_s \frac{q_s}{m_s}\rho_s\bE$ term on the left hand side. The left hand side of equation \eqref{eq:qnEeqn} results in a symmetric, positive definite matrix after discretization, making it straightforward to solve.
%%%%%%%%%%%%%%%%%%%%%%%%%%%%%%%%%%%%%%%%%%%%%%%%%%%%%%%%%

%%%%%%%%%%%%%%%%%%%%%%%%%%%%%%%%%%%%%%%%%%%%%%%%%%%%%%%%%
\section{Spatial discretization of fields}\label{sec:fielddiscretization}

\begin{figure}
    \centering
    \includegraphics[width=0.95\linewidth]{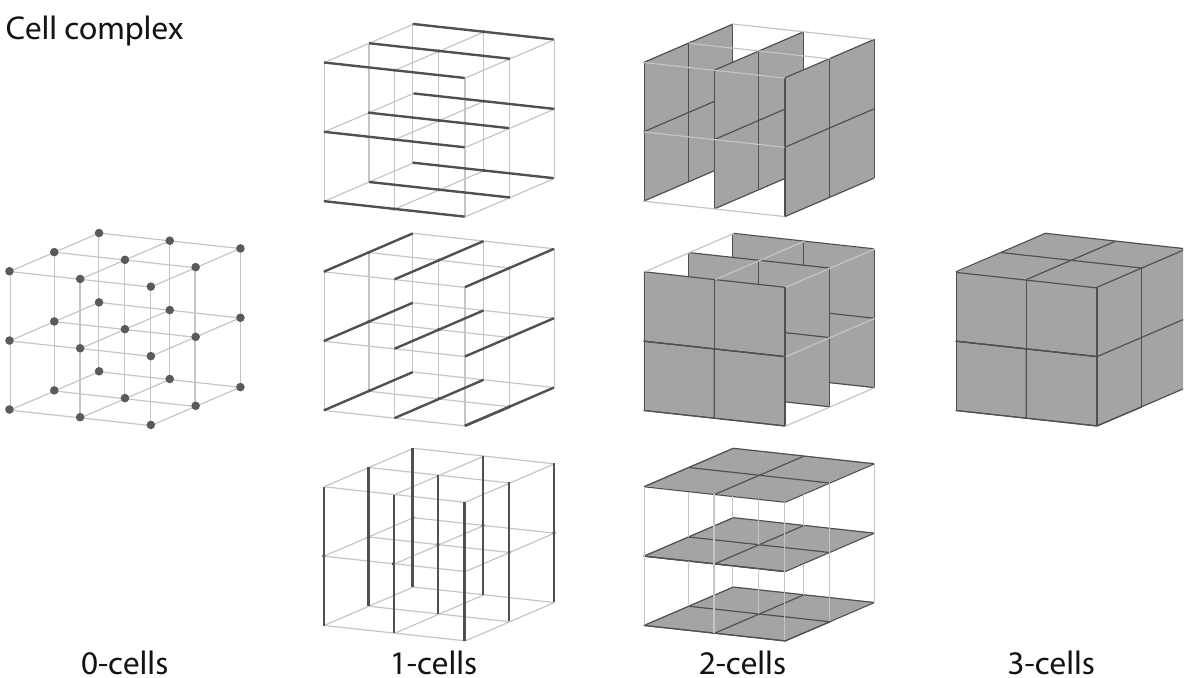}
    \caption{Spaces for defining discrete variables on a primal Cartesian grid.}
    \label{fig:CellComplexes}
\end{figure}

We use a dual-grid approach for the spatial discretization of our model, wherein the dual or adjoint grid vertices are the barycenters of the primal grid cells. The numerical values for the field variables are located on spaces based on points, edges, faces and volumes of hexahedral (3D) or quadrilateral (2D) cells. This discretization, based on mimetic finite differences, has already been used in \texttt{GEMPICX} for discretizing the fully kinetic \citep{kormann2024} and the drift-kinetic \citep{meng2025} Vlasov-Maxwell models. The nodal, edge, face and volume spaces on the primal grids are denoted by $\cC_0$, $\cC_1$, $\cC_2$ and $\cC_3$, respectively. Similarly, those on the dual grids are denoted by $\tilde \cC_0$, $\tilde \cC_1$, $\tilde \cC_2$ and $\tilde \cC_3$, respectively. A duality exists between the primal and dual spaces. The nodes, edges, faces and cell volumes on the primal grid are each uniquely associated with the cell volumes, faces, edges and nodes on the dual grid, respectively. On the primal grid, the electric field $\bE$ and the vector potential $\bA$ are defined as edge-integrals and the magnetic field $\bB$ is defined as face-integrals. On the dual grid, the magnetic field intensity $\bH$ is defined as edge-integrals, the current density $\bJ$ and electric field intensity $\bD$ are defined as face-integrals, and the charge density $\rho$ is defined as volume-integrals. Such an approach allows for exact preservation of certain discretized quantities. For example, the discrete divergence of $\bJ$ becomes a volume-integral on the dual grid, which is the same space where the charge density $\rho$ is defined. These spaces are shown in Figure \ref{fig:CellComplexes} for the primal grid.

\subsection{Reduction operators}\label{sec:reduction}
The reduction operators, also called restriction operators, relate a continuous field to its discretized counterparts defined on nodes, edges, faces and volumes. On the primal grid, these are:
\begin{itemize}
	\item $\cR_0$ relates a scalar field to its values at primal grid nodes:
    \begin{align}
    \cR_0(\psi)_{i,j,k} = \psi(x_i, y_j, z_k) = {\bm{\uppsi}}_{i,j,k}.
    \end{align}
    Here, $\bm{\uppsi} \in \cC_0$.

    \item $\cR_1$ relates a vector field to its integrals on primal grid edges:
    \begin{gather}
    \cR_1^x(E_x)_{i,j,k} = \int_{x_i}^{x_{i+1}} E_x(x, y_j, z_k) \dd x = {\arr{E}}_{i+1/2,j,k} = {\arr{E}}^x_{i,j,k},\\
    \cR_1^y(E_y)_{i,j,k} = \int_{y_j}^{y_{j+1}} E_y(x_i, y, z_k) \dd y = {\arr{E}}_{i,j+1/2,k} = {\arr{E}}^y_{i,j,k},\\
    \cR_1^z(E_z)_{i,j,k} = \int_{z_k}^{z_{k+1}} E_z(x_i, y_j, z) \dd z = {\arr{E}}_{i,j,k+1/2} = {\arr{E}}^z_{i,j,k}.
    \end{gather}
    Here, $\cR_1(\bE) = (\cR_1^x(E_x), \cR_1^y(E_y), \cR_1^z(E_z)) = ({\arr{E}}^x,{\arr{E}}^y,{\arr{E}}^z)$ and $\arr{E} = (\arr{E}^x, \arr{E}^y, \arr{E}^z) \in \cC_1$.

    \item $\cR_2$ relates a vector field to its integrals on primal grid faces:
    \begin{gather}
    \cR_2^x(B_x)_{i,j,k} = \int_{z_k}^{z_{k+1}} \int_{y_j}^{y_{j+1}} B_x(x_i, y, z) \dd y\dd z = {\arr{B}}_{i,j+1/2,k+1/2} = {\arr{B}}^x_{i,j,k},\\
    \cR_2^y(B_y)_{i,j,k} = \int_{z_k}^{z_{k+1}} \int_{x_i}^{x_{i+1}} B_y(x, y_j, z) \dd x\dd z = {\arr{B}}_{i+1/2,j,k+1/2} = {\arr{B}}^y_{i,j,k},\\
    \cR_2^z(B_z)_{i,j,k} = \int_{y_j}^{y_{j+1}} \int_{x_i}^{x_{i+1}} B_z(x, y, z_k) \dd x\dd y = {\arr{B}}_{i+1/2,j+1/2,k} = {\arr{B}}^z_{i,j,k}.
    \end{gather}
    Here, $\cR_2(\bB) = (\cR_2^x(B_x), \cR_2^y(B_y), \cR_2^z(B_z)) = ({\arr{B}}^x,{\arr{B}}^y,{\arr{B}}^z)$ and $\arr{B} = (\arr{B}^x, \arr{B}^y, \arr{B}^z) \in \cC_2$.

	\item $\cR_3$ relates a scalar field to its volume integrals over primal grid cells:
    \begin{gather}
    \cR_3(\rho)_{i,j,k} = \int_{z_k}^{z_{k+1}} \int_{y_j}^{y_{j+1}} \int_{x_i}^{x_{i+1}} \rho(x, y, z) \dd x\dd y\dd z = {\bm{\uprho}}_{i,j,k}.
    \end{gather}
    Here, $\bm{\uprho} \in \cC_3$.
\end{itemize}
The reduction/restriction operators on the dual mesh are similarly defined as follows:
\begin{itemize}
	\item $\tilde\cR_0$ relates a scalar field to its values at dual grid nodes:
    \begin{gather}
    \tilde\cR_0(\phi)_{i,j,k} = \phi(x_{i+1/2}, y_{j+1/2}, z_{k+1/2}) = \tilde{\bm{\upphi}}_{i+1/2,j+1/2,k+1/2}.
    \end{gather}
    Here, $\tilde{\bm{\upphi}} \in \tilde\cC_0$.

    \item $\tilde\cR_1$ relates a vector field to its integrals on dual grid edges:
    \begin{gather}
    \tilde\cR_1^x(H_x)_{i,j,k} = \int_{x_{i-1/2}}^{x_{i+1/2}} H_x(x, y_{j+1/2}, z_{k+1/2}) \dd x = \tilde{\arr{H}}_{i,j+1/2,k+1/2} = \tilde{\arr{H}}^x_{i,j,k},\\
    \tilde\cR_1^y(H_y)_{i,j,k} = \int_{y_{j-1/2}}^{y_{j+1/2}} H_y(x_{i+1/2}, y, z_{k+1/2}) \dd y = \tilde{\arr{H}}_{i+1/2,j,k+1/2} = \tilde{\arr{H}}^y_{i,j,k},\\
    \tilde\cR_1^z(H_z)_{i,j,k} = \int_{z_{k-1/2}}^{z_{k+1/2}} H_z(x_{i+1/2}, y_{j+1/2}, z) \dd z = \tilde{\arr{H}}_{i+1/2,j+1/2,k} = \tilde{\arr{H}}^z_{i,j,k}.
    \end{gather}
    Here, $\tilde\cR_1(\bH) = (\tilde\cR_1^x(H_x), \tilde\cR_1^y(H_y), \tilde\cR_1^z(H_z)) = (\tilde{\arr{H}}^x, \tilde{\arr{H}}^y, \tilde{\arr{H}}^z)$ and $\tilde{\arr{H}} = (\tilde{\arr{H}}^x, \tilde{\arr{H}}^y, \tilde{\arr{H}}^z) \in \tilde\cC_1$.

    \item $\tilde\cR_2$ relates a vector field to its integrals on dual grid faces:
    \begin{gather}
    \tilde\cR_2^x(D_x)_{i,j,k} = \int_{z_{k-1/2}}^{z_{k+1/2}} \int_{y_{j-1/2}}^{y_{j+1/2}} D_x(x_{i+1/2}, y, z) \dd y\dd z = \tilde{\arr{D}}_{i+1/2,j,k} = \tilde{\arr{D}}^x_{i,j,k},\\
    \tilde\cR_2^y(D_y)_{i,j,k} = \int_{z_{k-1/2}}^{z_{k+1/2}} \int_{x_{i-1/2}}^{x_{i+1/2}} D_y(x, y_{j+1/2}, z) \dd x\dd z = \tilde{\arr{D}}_{i,j+1/2,k} = \tilde{\arr{D}}^y_{i,j,k},\\
    \tilde\cR_2^z(D_z)_{i,j,k} = \int_{y_{j-1/2}}^{y_{j+1/2}} \int_{x_{i-1/2}}^{x_{i+1/2}} D_z(x, y, z_{k+1/2}) \dd x\dd y = \tilde{\arr{D}}_{i,j,k+1/2} = \tilde{\arr{D}}^z_{i,j,k}.
    \end{gather}
    Here, $\tilde\cR_2(\bD) = (\tilde\cR_2^x(D_x), \tilde\cR_2^y(D_y), \tilde\cR_2^z(D_z)) = (\tilde{\arr{D}}^x, \tilde{\arr{D}}^y, \tilde{\arr{D}}^z)$ and $\tilde{\arr{D}} = (\tilde{\arr{D}}^x, \tilde{\arr{D}}^y, \tilde{\arr{D}}^z) \in \tilde\cC_2$.

	\item $\tilde\cR_3$ relates a scalar field to its volume integrals over dual grid cells:
    \begin{gather}
    \tilde\cR_3(\rho)_{i,j,k} = \int_{z_{k-1/2}}^{z_{k+1/2}} \int_{y_{j-1/2}}^{y_{j+1/2}} \int_{x_{i-1/2}}^{x_{i+1/2}} \rho(x, y, z) \dd x\dd y\dd z = \tilde{\bm{\uprho}}_{i,j,k}.
    \end{gather}
    Here, $\tilde{\bm{\uprho}} \in \tilde\cC_3$.
\end{itemize}

\subsection{Hodge operators}\label{sec:hodge}
Field variables can be projected from primal grid spaces to their dual counterparts or vice versa using Hodge operators. Hodge operators of an arbitrary order of accuracy can be obtained using the method prescribed by \citet{kormann2024}. In the current work, only second-order Hodge projections have been used and this simply constitutes the use of scaling factors to relate variables between primal and dual spaces. Consider arbitrary variables $\bK$, $\bL$, $\bM$ and $\bN$, defined on primal nodes, edges, faces and cell volumes, respectively. Their corresponding dual mappings, defined on dual cell volumes, faces, edges and nodes are denoted as $\tilde\bK$, $\tilde\bL$, $\tilde\bM$ and $\tilde\bN$, respectively. These would be given by:
\begin{gather}
  \tilde{\arr{K}} = \Delta x\, \Delta y\, \Delta z \arr{K}, \notag \\
  \tilde{\arr{L}}^x = \frac{\Delta y\, \Delta z}{\Delta x} \arr{L}^x, \quad
  \tilde{\arr{L}}^y = \frac{\Delta x\, \Delta z}{\Delta y} \arr{L}^y, \quad
  \tilde{\arr{L}}^z = \frac{\Delta x\, \Delta y}{\Delta z} \arr{L}^z, \notag \\
  \tilde{\arr{M}}^x = \frac{\Delta x}{\Delta y\, \Delta z} \arr{M}^x, \quad
  \tilde{\arr{M}}^y = \frac{\Delta y}{\Delta x\, \Delta z} \arr{M}^y, \quad
  \tilde{\arr{M}}^z = \frac{\Delta z}{\Delta x\, \Delta y} \arr{M}^z, \notag \\
  \tilde{\arr{N}} = \frac{1}{\Delta x\, \Delta y\, \Delta z} \arr{N}.
\end{gather}
We can therefore define Hodge operators $\matH_0$, $\matH_1$, $\matH_2$, and $\matH_3$, and their corresponding inverses $\tilde \matH_3$, $\tilde \matH_2$, $\tilde \matH_1$, and $\tilde \matH_0$, respectively:
\begin{gather}
  \tilde{\arr{K}} = \matH_0 \arr{K},\;\hspace{0.5cm} \arr{K} = \tilde \matH_3 \tilde{\arr{K}}, \notag \\
  \tilde{\arr{L}} = \matH_1 \arr{L},\;\hspace{0.5cm} \arr{L} = \tilde \matH_2 \tilde{\arr{L}}, \notag \\
  \tilde{\arr{M}} = \matH_2 \arr{M},\;\hspace{0.5cm} \arr{M} = \tilde \matH_1 \tilde{\arr{M}}, \notag \\
  \tilde{\arr{N}} = \matH_3 \arr{N},\;\hspace{0.5cm} \arr{N} = \tilde \matH_0 \tilde{\arr{N}}.
\end{gather}
Accordingly, the electric and magnetic fields can be related to their intensities as follows:
\begin{gather}\label{eq:hodge_relns}
  \tilde{\arr{D}} = \matH_1 \arr{E},\;\hspace{0.5cm} \arr{E} = \tilde \matH_2 \tilde{\arr{D}}, \notag \\
  \tilde{\arr{H}} = \matH_2 \arr{B},\;\hspace{0.5cm} \arr{B} = \tilde \matH_1 \tilde{\arr{H}}.
\end{gather}
\nmn{The second order expressions of the various Hodge operators are simply the appropriate scaling factors multiplied by the identity matrix. This is equivalent to the classical Yee scheme. While the current work only uses second-order Hodge operators, the mimetic framework admits Hodge constructions of arbitrary order \citep{kormann2024}. In the context of the full Vlasov-Maxwell system, higher-order Hodges have been shown to improve the accuracy of discrete differential operators and numerical dispersion properties, particularly at higher wavenumbers. To the best of our knowledge, however, such higher-order Hodge operators have not yet been investigated specifically within a geometric quasineutral PIC framework. Since the quasineutral approximation removes high-frequency electromagnetic modes while retaining low-frequency plasma dynamics, a dedicated assessment of the benefits of higher-order Hodges for quasineutral simulations remains to be seen.}

\subsection{Discrete scalar products}\label{sec:scalarproducts}
We can now also define discrete scalar products between primal scalar/vector fields and their dual counterparts to estimate $L^2$ inner products. For the variables $\bK$, $\bL$, $\bM$ and $\bN$ defined above, these discrete scalar products are as follows:
\begin{equation}
  \arr{K} \cdot \tilde{\arr{K}} = \sum_{i,j,k} \arr{K}_{i,j,k} \tilde{\arr{K}}_{i,j,k},
\end{equation}
\begin{gather}
  \arr{L} \cdot \tilde{\arr{L}} 
  = \arr{L}^x \cdot \tilde{\arr{L}}^x 
   + \arr{L}^y \cdot \tilde{\arr{L}}^y 
   + \arr{L}^z \cdot \tilde{\arr{L}}^z \notag \\
  = \sum_{i,j,k} \big(
     \arr{L}_{i+1/2,j,k} \tilde{\arr{L}}_{i+1/2,j,k}
   + \arr{L}_{i,j+1/2,k} \tilde{\arr{L}}_{i,j+1/2,k}
   + \arr{L}_{i,j,k+1/2} \tilde{\arr{L}}_{i,j,k+1/2}
   \big),
\end{gather}
\begin{gather}
  \arr{M} \cdot \tilde{\arr{M}} 
  = \arr{M}^x \cdot \tilde{\arr{M}}^x 
   + \arr{M}^y \cdot \tilde{\arr{M}}^y 
   + \arr{M}^z \cdot \tilde{\arr{M}}^z \notag \\
  = \sum_{i,j,k} \big(
     \arr{M}_{i,j+1/2,k+1/2} \tilde{\arr{M}}_{i,j+1/2,k+1/2}
   + \arr{M}_{i+1/2,j,k+1/2} \tilde{\arr{M}}_{i+1/2,j,k+1/2} \notag \\
  \hphantom{=\sum_{i,j,k} \big(}
   + \arr{M}_{i+1/2,j+1/2,k} \tilde{\arr{M}}_{i+1/2,j+1/2,k}
   \big),
\end{gather}

\begin{equation}
  \arr{N} \cdot \tilde{\arr{N}} = \sum_{i,j,k} \arr{N}_{i+1/2,j+1/2,k+1/2} \tilde{\arr{N}}_{i+1/2,j+1/2,k+1/2}.
\end{equation}

\subsection{Discrete gradient, curl and divergence}\label{sec:discretederivatives}
In order to define the discrete gradient, curl and divergence operators, we first define a one-dimensional discrete derivative or difference operator
\begin{equation*}
	\mathbb{d}_{M_1} = \begin{pmatrix}
   -1 & 1 & 0 & \ldots & 0 \\
   0 & -1 & 1 & 0  & \\
   \vdots & & \ddots & \ddots & \\
   0 & & & -1 & 1 \\
   1 & 0 & \ldots & 0 & -1
   \end{pmatrix} \in \mathbb{R}^{M_1 \times M_1}.
\end{equation*}
We can represent a $\mathbb{d}$ matrix and an identity matrix of size $N$ as $\mathbb{d}_N$ and and $\matI_N$, respectively. Consider a grid with $N_1$, $N_2$, $N_3$ points in the $x$-, $y$- and $z$- directions, respectively, such that $N = N_1 \times N_2 \times N_3$. For such a grid, the discrete gradient, curl, and divergence operators on the primal grid can be denoted by $\mathbb{G}$, $\mathbb{C}$ and $\mathbb{D}$. These are built using Kronecker products of the $\mathbb{d}$ and $\matI$ matrices of appropriate sizes as follows

\begin{equation} \label{hD0}
\matG = \begin{pmatrix}
\matd_{N_1} \otimes \matI_{N_2} \otimes \matI_{N_3} \\
\matI_{N_1} \otimes \matd_{N_2} \otimes \matI_{N_3} \\
\matI_{N_1} \otimes \matI_{N_2} \otimes \matd_{N_3}
\end{pmatrix},
\end{equation}

\begin{equation}\label{hD12}
\matC = \begin{pmatrix}
\matO_{N} & - \matI_{N_1} \otimes \matI_{N_2} \otimes \matd_{N_3}  & \matI_{N_1} \otimes \matd_{N_2} \otimes \matI_{N_3}\\
\matI_{N_1} \otimes \matI_{N_2} \otimes \matd_{N_3}  & \matO_{N} &  -\matd_{N_1} \otimes \matI_{N_2} \otimes \matI_{N_3}\\
-\matI_{N_1} \otimes \matd_{N_2} \otimes \matI_{N_3}  & \matd_{N_1} \otimes \matI_{N_2} \otimes \matI_{N_3} & \matO_{N} \\
\end{pmatrix},
\end{equation}

\begin{equation} \label{hD3}
\matD = \begin{pmatrix}
\matd_{N_1} \otimes \matI_{N_2} \otimes \matI_{N_3} & \matI_{N_1} \otimes \matd_{N_2} \otimes \matI_{N_3} &  \matI_{N_1} \otimes \matI_{N_2} \otimes \matd_{N_3} \\
\end{pmatrix}.
\end{equation}
The adjoint operators of these operators on the dual grid are given by
\begin{gather}
\tilde \matG = - \matD^T, \\
\tilde \matC = \matC^T, \\
\tilde \matD = - \matG^T.
\end{gather}
Due to our degrees of freedom being defined on discrete spaces associated with the de Rham complex, the discrete gradient, curl and divergence operators and their dual operators are always exact.
%%%%%%%%%%%%%%%%%%%%%%%%%%%%%%%%%%%%%%%%%%%%%%%%%%%%%%%%%

%%%%%%%%%%%%%%%%%%%%%%%%%%%%%%%%%%%%%%%%%%%%%%%%%%%%%%%%%
\section{Particle in Cell (PIC) discretization}\label{sec:particlediscretization}

Let us now consider the particle discretization of the Vlasov equation, where we write
\begin{equation} \label{eq:fpart}
	f_s(t,\bx,\bv) = \sum_{p=1}^{N_s} w_p \delta(\bx -\bx_p(t))\delta(\bv -\bv_p(t)),
\end{equation}
where $w_p$ is the particle weight, $N_s$ is the number of particles of species $s$ and $\delta$ is the Dirac delta function. This function is clearly not continuous and smooth. However, in order to use particle information to calculate fields such as $\rho$, $\bJ$ and $\mathbb{S}$ in their discretized forms, we would require a smoothing kernel, typically a spline. Such a spline would effectively spread out the influence of the particle's mass and charge over a volume centered at the particle location, rather than keep it concentrated at a single point. These splines would also be used to calculate electric and magnetic fields at particle locations using the discretized forms of the fields. As each particle carries its own charge $q_p$, mass $m_p$ and weight $w_p$, we shall not distinguish the species in the discrete sum and $\sum_p$ will denote in the sequel the sum over all particles of all species. Each particle then obeys equations \eqref{eq:Xdots} and \eqref{eq:Vdots}. The splines used in this work are based on the well-known cardinal B-splines. Fundamental cardinal B-splines of an arbitrary degree $d$ centered at $x = 0$ can be defined recursively using the convolution

\begin{equation}
S^{(d)}(x) = S^{(0)} * S^{(d-1)}(x) = \int_{-1/2}^{1/2} S^{(d-1)}(x - x') \, dx',
\end{equation}
with
\begin{equation}
S^{(0)}(x) =
\begin{cases}
1 & \text{if } -\frac{1}{2} \le x \le \frac{1}{2} \\
0 & \text{otherwise}
\end{cases}.
\end{equation}
Cardinal B-splines have the property
\begin{equation}
\int_{-\infty}^{\infty} S^{(d)}(x) \, dx = 1.
\end{equation}
The coupling of particles and fields involves two kinds of B-splines: node splines and cell splines. These splines are built using the cardinal B-spline described above.

These splines are key to evaluation of $\bE$, $\bB$ fields at particle positions and evaluation of field variables like charge density $\rho$ and current density $\bJ$, as defined in their respective spaces.
Consider the mesh size in the $x$-direction to be denoted as $h$. Node splines are centered on grid nodes, and are defined as
\begin{equation}
S_i^{n}(x_p) = S^d((x_p - x_i)/h).
\end{equation}
This spline is supported on the interval $[x_i - (d+1)h/2, x_i + (d+1)h/2]$, and its values are independent of the grid size.
Cell splines are centered on cell midpoints, and are defined as
\begin{equation}
S_i^{c}(x_p) = \frac{1}{h}S^{d-1}\left(\frac{x_p - x_{i+1/2}}{h}\right).
\end{equation}
This spline is supported on the interval $[x_{i+1/2} - dh/2, x_{i+1/2} + dh/2]$, and its volume-integrals are independent of the grid size. Node and cell splines for the $y$- and $z$- directions can be similarly defined given the respective mesh sizes in those directions.

The charge density $\rho$ is defined as volume-integrals on the dual grid i.e. $\tilde{\bm{\uprho}} = \tilde\cR_3(\rho)  \in \tilde\cC_3$. This is calculated as
\begin{equation}
\tilde{\bm{\uprho}}(t)_{i,j,k} = \tilde\cR_3(\rho(t))_{i,j,k} = \sum_p q_p w_p S_i^{n}(x_p(t)) S_j^{n}(y_p(t)) S_k^{n}(z_p(t)).
\end{equation}
The current density $\bJ$ is defined as face integrals on the dual grid i.e. $\tilde{\arr{J}} = \tilde\cR_2(\bJ)  \in \tilde\cC_2$. This is calculated as
\begin{gather}
\tilde{\arr{J}}^x(t)_{i+1/2,j,k} = \tilde\cR_2(J^x(t))_{i+1/2,j,k} = \sum_p q_p w_p v_{p,x} S_i^{c}(x_p(t)) S_j^{n}(y_p(t)) S_k^{n}(z_p(t)),\label{eq:depositjx} \\
\tilde{\arr{J}}^y(t)_{i,j+1/2,k} = \tilde\cR_2(J^y(t))_{i,j+1/2,k} = \sum_p q_p w_p v_{p,y} S_i^{n}(x_p(t)) S_j^{c}(y_p(t)) S_k^{n}(z_p(t)), \label{eq:depositjy}\\
\tilde{\arr{J}}^z(t)_{i,j,k+1/2} = \tilde\cR_2(J^z(t))_{i,j,k+1/2} = \sum_p q_p w_p v_{p,z} S_i^{n}(x_p(t)) S_j^{n}(y_p(t)) S_k^{c}(z_p(t)). \label{eq:depositjz}
\end{gather}

The components of the divergence of the stress tensor $\nabla \cdot \mathbb{S}$ are also defined as face integrals on the dual grid. The matrix $\mathbb{S}$ can be written in terms of its components as
\begin{equation*}
	\mathbb{S} = \begin{pmatrix}
   \mathbb{S}^x \\
   \mathbb{S}^y \\
   \mathbb{S}^z
   \end{pmatrix} = 
   \begin{pmatrix}
\mathbb{S}^{xx} & \mathbb{S}^{xy} & \mathbb{S}^{xz} \\
\mathbb{S}^{yx} & \mathbb{S}^{yy} & \mathbb{S}^{yz} \\
\mathbb{S}^{zx} & \mathbb{S}^{zy} & \mathbb{S}^{zz} \\
\end{pmatrix}.
\end{equation*}
Similarly, the vector $\nabla \cdot \mathbb{S}$ can be written as
\begin{equation*}
	\nabla \cdot \mathbb{S} = \begin{pmatrix}
   (\nabla \cdot \mathbb{S})^x \\
   (\nabla \cdot \mathbb{S})^y \\
   (\nabla \cdot \mathbb{S})^z
   \end{pmatrix} =
   \begin{pmatrix}
\partial_x(\mathbb{S}^{xx}) + \partial_y(\mathbb{S}^{xy}) + \partial_z(\mathbb{S}^{xz}) \\
\partial_x(\mathbb{S}^{yx}) + \partial_y(\mathbb{S}^{yy}) + \partial_z(\mathbb{S}^{yz}) \\
\partial_x(\mathbb{S}^{zx}) + \partial_y(\mathbb{S}^{zy}) + \partial_z(\mathbb{S}^{zz}) \\
\end{pmatrix}.
\end{equation*}
If we denote $\nabla \cdot \mathbb{S}$ as $\bQ$ and its discrete dual face equivalent as $\tilde{\arr{Q}}$, we have $\tilde{\arr{Q}} = \tilde\cR_2(\bQ) \in \tilde\cC_2$. We denote the dual equivalent of the term $\partial_y(\mathbb{S}^{xy})$ as $\tilde{\arr{Q}}^{xy}$ and so on for other terms. Thus, $\tilde{\arr{Q}}$ can be written as
\begin{equation*}
	\tilde{\arr{Q}} = \begin{pmatrix}
   \tilde{\arr{Q}}^x \\
   \tilde{\arr{Q}}^y \\
    \tilde{\arr{Q}}^z
   \end{pmatrix} =
   \begin{pmatrix}
 \tilde{\arr{Q}}^{xx} + \tilde{\arr{Q}}^{xy} + \tilde{\arr{Q}}^{xz} \\
\tilde{\arr{Q}}^{yx} + \tilde{\arr{Q}}^{yy} + \tilde{\arr{Q}}^{yz} \\
\tilde{\arr{Q}}^{zx} + \tilde{\arr{Q}}^{zy} + \tilde{\arr{Q}}^{zz} \\
\end{pmatrix}.
\end{equation*}
The individual components of $\tilde{\arr{Q}}$ are calculated as
\begin{gather}
\tilde{\arr{Q}}^{xx}(t)_{i+1/2,j,k} = \sum_p q_p w_p v_{p,x} v_{p,x} \fracp{S_i^{c}(x_p(t))}{x_p(t)} S_j^{n}(y_p(t)) S_k^{n}(z_p(t)), \\
\tilde{\arr{Q}}^{xy}(t)_{i+1/2,j,k} = \sum_p q_p w_p v_{p,x} v_{p,y} S_i^{c}(x_p(t)) \fracp{S_j^{n}(y_p(t))}{y_p(t)} S_k^{n}(z_p(t)), \\
\tilde{\arr{Q}}^{xz}(t)_{i+1/2,j,k} = \sum_p q_p w_p v_{p,x} v_{p,z} S_i^{c}(x_p(t)) S_j^{n}(y_p(t)) \fracp{S_k^{n}(z_p(t))}{z_p(t)}, \\
\tilde{\arr{Q}}^{yx}(t)_{i,j+1/2,k} = \sum_p q_p w_p v_{p,y} v_{p,x} \fracp{S_i^{n}(x_p(t))}{x_p(t)} S_j^{c}(y_p(t)) S_k^{n}(z_p(t)), \\
\tilde{\arr{Q}}^{yy}(t)_{i,j+1/2,k} = \sum_p q_p w_p v_{p,y} v_{p,y} S_i^{n}(x_p(t)) \fracp{S_j^{c}(y_p(t))}{y_p(t)} S_k^{n}(z_p(t)), \\
\tilde{\arr{Q}}^{yz}(t)_{i,j+1/2,k} = \sum_p q_p w_p v_{p,y} v_{p,z} S_i^{n}(x_p(t)) S_j^{c}(y_p(t)) \fracp{S_k^{n}(z_p(t))}{z_p(t)}, \\
\tilde{\arr{Q}}^{zx}(t)_{i,j,k+1/2} = \sum_p q_p w_p v_{p,z} v_{p,x} \fracp{S_i^{n}(x_p(t))}{x_p(t)} S_j^{n}(y_p(t)) S_k^{c}(z_p(t)), \\
\tilde{\arr{Q}}^{zy}(t)_{i,j,k+1/2} = \sum_p q_p w_p v_{p,z} v_{p,y} S_i^{n}(x_p(t)) \fracp{S_j^{n}(y_p(t))}{y_p(t)} S_k^{c}(z_p(t)), \\
\tilde{\arr{Q}}^{zz}(t)_{i,j,k+1/2} = \sum_p q_p w_p v_{p,z} v_{p,z} S_i^{n}(x_p(t)) S_j^{n}(y_p(t)) \fracp{S_k^{c}(z_p(t))}{z_p(t)}.
\end{gather}

The values of the electric and magnetic fields at the particle locations, i.e. $\bE^S = (E_x^S, E_y^S, E_z^S)$ and $B^S = (B_x^S, B_y^S, B_z^S)$, come from an integration of the fields over the particle shape functions. The exact expressions given here can be derived from the discrete Lagrangian presented in the next section. These are used to update particle velocities. These are also calculated at particle positions using node and cell splines. These fields are calculated in terms of the edge-integrals $\arr{E}$ and face-integrals $\arr{B}$.
\begin{gather}
E_x^S(\bx_p) = \sum_{i,j,k} {\arr{E}}_{i+1/2,j,k} S_i^{c}(x_p) S_j^{n}(y_p) S_k^{n}(z_p), \\
E_y^S(\bx_p) = \sum_{i,j,k} {\arr{E}}_{i,j+1/2,k} S_i^{n}(x_p) S_j^{c}(y_p) S_k^{n}(z_p), \\
E_z^S(\bx_p) = \sum_{i,j,k} {\arr{E}}_{i,j,k+1/2} S_i^{n}(x_p) S_j^{n}(y_p) S_k^{c}(z_p).
\end{gather}

\begin{gather}
B_x^S(\bx_p) = \sum_{i,j,k} {\arr{B}}_{i,j+1/2,k+1/2} S_i^{n}(x_p) S_j^{c}(y_p) S_k^{c}(z_p), \\
B_y^S(\bx_p) = \sum_{i,j,k} {\arr{B}}_{i+1/2,j,k+1/2} S_i^{c}(x_p) S_j^{n}(y_p) S_k^{c}(z_p), \\
B_z^S(\bx_p) = \sum_{i,j,k} {\arr{B}}_{i+1/2,j+1/2,k} S_i^{c}(x_p) S_j^{c}(y_p) S_k^{n}(z_p).
\end{gather}
The dual face equivalents of the Lorentz force $\bJ \times \bB$ can be obtained by simply replacing the components of $\bv_p$ in equations \eqref{eq:depositjx}--\eqref{eq:depositjz} with components of $\bv_p\times \bB^S(\bx_p)$.

%%%%%%%%%%%%%%%%%%%%%%%%%%%%%%%%%%%%%%%%%%%%%%%%%%%%%%%%%

%%%%%%%%%%%%%%%%%%%%%%%%%%%%%%%%%%%%%%%%%%%%%%%%%%%%%%%%%
\section{Semi-discretized governing equations}\label{sec:discreteqnvmmodel}
In this section, we use the semi-discrete action principle to obtain the semi-discrete governing equations. The discretization is only in space, while all quantities are still varying continuously with time. The semi-discrete electric field equation is also derived. Conservation of energy for our proposed semi-discretization is mathematically proven.

\subsection{Semi-discrete action principle}\label{sec:discreteaction}
In this framework, in the same way as in \citet{kormann2024}, we can write a discrete action principle from which our model will follow:
\begin{gather}
	\mathcal{L}_{h} = \sum_{p=1}^{N_p} \left[w_p (m_p \bv_p + q_p \arr{A}) \cdot \tilde{\mathcal{R}}_2  \left( \dot{\bx}_p S(\bx-\bx_p)\right) - \frac 12 m_p \bv_p^2 - q_p \arr{\upphi} \cdot \tilde{\mathcal{R}}_3(S(\bx-\bx_p)) \right] \notag \\ - \frac{1}{2\mu_0} (\mathbb{C}\arrA) \cdot (\mathbb{H}_1\mathbb{C}\arrA). \label{eq:discL}
\end{gather}
The discretized Euler-Lagrange equations for the particles, obtained by taking the variations with respect to $\bx_p$ and $\bv_p$ are given by
\begin{gather}
	\fract{\bx_p}{t} =\bv_p, \label{eq:Xdotsdisc} \\
	\fract{\bv_p}{t} = \frac {q_p}{m_p}\left(\bE^S(t,\bx_p) + \bv_p \times \bB^S(t,\bx_p)\right).\label{eq:Vdotsdisc}
\end{gather}
In our discrete setting, the electromagnetic fields and the potentials, all defined on the primal grid, are related by
\begin{equation} \label{eq:disc_field_defns}
    \arrE = -\frac{\dd \arrA}{\dd t} - \mathbb{G}\arr{\upphi}, ~~~~ \arrB = \mathbb{C}\arrA.
\end{equation}
This immediately yields the discrete Faraday equation
\begin{equation} \label{eq:faraday_disc}
    \fract{\arrB}{t} + \mathbb{C}\arrE = 0.
\end{equation}
Similarly, the definition of the magnetic field, coupled with the discretization of the magnetic field on primal grid faces, allows us to write the discretized Gauss law of magnetism
\begin{equation} \label{eq:gaussB_disc}
\mathbb{D}\arrB = 0.
\end{equation}
Taking the variations of $\mathcal{L}_{h}$ with respect to $\arrA$, we find Amp\`ere's law
\begin{equation} \label{eq:ampere_disc}
    \mathbb{C}^\top\mathbb{H}_2\arrB = \mathbb{C}^\top\mathbb{H}_2 \mathbb{C}\arrA = \mu_0\tilde{\arr{J}}, ~~~ \mbox{where } \tilde{\arr{J}}(t) = \sum_p w_p q_p \tilde{\mathcal{R}}_2  \left( \bv_p(t) S(\bx-\bx_p(t)\right).
\end{equation}
Taking the variations with respect to $\arr{\upphi}$ leads to the discretized quasineutrality condition
\begin{equation} \label{eq:quasineutrality_disc}
    \tilde{\bm{\uprho}} = 0, ~~~ \mbox{where } \tilde{\bm{\uprho}} = \sum_p w_p q_p \tilde{\mathcal{R}}_3  \left(S(\bx-\bx_p(t)\right).
\end{equation}
Just like the continuous counterpart, this results in the discrete divergence of the current density being 0 i.e.
\begin{equation} \label{eq:divJ_disc}
\tilde{\mathbb{D}}\tilde{\arrJ} = 0.
\end{equation}
While equation \eqref{eq:qnrho} is satisfied at the continuous level of the quasineutral model, at the discrete level equation \eqref{eq:quasineutrality_disc} is not satisfied to machine precision. The discrete charge density volume integrals are obtained by depositing a finite number of particles onto the grid, introducing statistical sampling noise. The resulting deviations are typically small, and since quasineutrality is embedded in the electric and magnetic field equations, they do not affect the consistency of the quasineutral limit. In PIC codes like the one presented here, these fluctuations decrease with increasing number of particles per cell, scaling approximately as $\mathcal{O}(1/\sqrt{N_p})$ \citep{birdsall2018plasma}. Instead, our numerical scheme ensures that the discrete quasineutrality divergence constraint given by equation \eqref{eq:divJ_disc} is satisfied up to machine precision at the discretized level, as explained later in Section \ref{sec:timestepping}. A divergence-free $\tilde{\arrJ}$ also leads to a divergence-free $\arrA$. This simplifies the discretized curl-curl operator in equation \eqref{eq:ampere_disc} to a discrete Laplace operator $\mathbb{L}$. Accordingly, the discretized counterpart of equation \eqref{eq:laplaceA} becomes the linear system
\begin{equation} \label{eq:laplaceA_disc}
(\mathbb{C}^\top\mathbb{H}_2 \mathbb{C} - \mathbb{H}_1 \mathbb{G} \tilde{\mathbb{H}}_3 \tilde{\mathbb{D}} \mathbb{H}_1)\arrA = -\mathbb{L} \arrA= \tilde{\arrJ}.
\end{equation}
After obtaining the divergence-free $\tilde{\arrJ}$, we use the above equation to obtain the divergence-free $\arrA$. The discretized magnetic field $\arrB$ is then obtained using $\arrB = \mathbb{C}\arrA$. This ensures that the discretized Amp\`ere's law, equation \eqref{eq:ampere_disc} is exactly satisfied by our structure-preserving scheme. Maintaining this consistency between the discretized current and Amp\`ere's law is essential to avoid unphysical behaviour in the numerical solution of the quasineutral model, as has been pointed out by \citet{camporeale2017electron}. We reiterate that our use of periodic boundary conditions here makes it easy to impose the discrete divergence-free  condition on $\arrA$ i.e. $\tilde{\mathbb{D}} \mathbb{H}_1 \arrA = 0$. This makes the discrete Laplace operator in equation \eqref{eq:laplaceA_disc} invertible to a constant, and also maintains the exactness of the discrete mimetic de Rham structure. Non-periodic boundaries would necessitate modifications in equation \eqref{eq:laplaceA_disc} to incorporate boundary effects.

\subsection{Semi-discrete electric field equation}\label{sec:discreteefieldeqn}
Equations \eqref{eq:faraday_disc}-\eqref{eq:ampere_disc} do not yield a direct way to update the electric field. For this reason, following \citet{Degond2017Asymptotic-Pres} and \citet{Degond2017AP-review}, the problem is reformulated to get an equation for the electric field. The time derivative of the discrete Amp\`ere's equation \eqref{eq:ampere_disc} using the particle current gives
\begin{gather}
	\mathbb{C}^\top\mathbb{H}_2\fract{\arrB}{t} = \mu_0\fract{\tilde\arrJ}{t} = \mu_0\sum_p q_p w_p \tilde{\mathcal{R}}_2\left(\fract{}{t}\left( \bv_p(t) S(\bx-\bx_p(t))\right)\right) \notag \\ =\mu_0\sum_p q_p w_p \tilde{\mathcal{R}}_2\left(\fract{\bv_p}{t} S(\bx-\bx_p) - \bv_p \fract{\bx_p}{t}\cdot \nabla S(\bx-\bx_p) \right) \notag \\
	=\mu_0\sum_p w_p \tilde{\mathcal{R}}_2\left(\frac{q_p^2}{m_p}(\bE^S(t,\bx_p) + \bv_p\times \bB^S(t,\bx_p) ) S(\bx-\bx_p) - q_p\bv_p \bv_p\cdot \nabla S(\bx-\bx_p) \right). \label{eq:genOhm}
\end{gather}
Then, taking the discrete curl of \eqref{eq:faraday_disc} yields the following equation for the electric field:
\begin{gather}
    \mathbb{C}^\top \mathbb{H}_2\mathbb{C}\arrE +
    \mu_0\sum_p \frac{w_p q_p^2}{m_p} \tilde{\mathcal{R}}_2\left(\bE^S(t,\bx_p) S(\bx-\bx_p(t))\right) = \notag \\ - \mu_0\sum_p w_p \tilde{\mathcal{R}}_2\left(\frac{q_p^2}{m_p}(\bv_p\times \bB^S(t,\bx_p) ) S(\bx-\bx_p) - q_p\bv_p \bv_p\cdot \nabla S(\bx-\bx_p) \right). \label{eq:efieldeqn_disc}
\end{gather}
This is a large, sparse, linear system that can be solved to obtain $\arrE$ in the numerical algorithm. The second term on the left hand side is a particle `mass' matrix, that requires looping over all the particles. This constitutes a significant portion of the overall computational cost. A cheaper alternative would be to instead use a diagonal matrix, $\mathbb{M}$, build from the discretized charge density contributions of the individual species, $\tilde{\bm{\uprho}}_s$, defined on the dual grid as volume-integrals. The resultant linear system using this approach would be
\begin{gather}
    \mathbb{C}^\top \mathbb{H}_2\mathbb{C}\arrE + \mathbb{M}\arrE = \notag \\ - \mu_0\sum_p w_p \tilde{\mathcal{R}}_2\left(\frac{q_p^2}{m_p}(\bv_p\times \bB^S(t,\bx_p) ) S(\bx-\bx_p) - q_p\bv_p \bv_p\cdot \nabla S(\bx-\bx_p) \right). \label{eq:efieldeqn_disc_rhofieldcorr}
\end{gather}
Here, the diagonal terms of $\mathbb{M}$, corresponding to the coefficients of the $x$-direction component of the discretized electric field, i.e. ${\arr{E}}_{i+1/2,j,k}$, or ${\arr{E}}^x_{i,j,k}$, are given by
\begin{equation}
    m_{i+1/2,j,k} = \sum_{s}\frac{q_s}{m_s}\bar{\rho}_{s,i+1/2,j,k} = \sum_{s}\frac{q_s}{m_s} \frac{\tilde{\bm{\uprho}}_{s,i,j,k} + \tilde{\bm{\uprho}}_{s,i+1,j,k}}{2\Delta V}.
\end{equation}
Here, $\bar{\rho}_{s,i+1/2,j,k}$ is an interpolated value of the charge density contribution of species $s$, obtained at the $x$-direction edge $(i+1/2,j,k)$ using the $\tilde{\bm{\uprho}}_s$ volume integrals from the dual cells $(i,j,k)$ and $(i+1,j,k)$, and the cell volume $\Delta V$. Similarly, the coefficients of ${\arr{E}}^y_{i,j,k}$ and ${\arr{E}}^z_{i,j,k}$ are given by
\begin{gather}
    m_{i,j+1/2,k} = \sum_{s}\frac{q_s}{m_s}\bar{\rho}_{s,i,j+1/2,k} = \sum_{s}\frac{q_s}{m_s} \frac{\tilde{\bm{\uprho}}_{s,i,j,k} + \tilde{\bm{\uprho}}_{s,i,j+1,k}}{2\Delta V}, \\
    m_{i,j,k+1/2} = \sum_{s}\frac{q_s}{m_s}\bar{\rho}_{s,i,j,k+1/2} = \sum_{s}\frac{q_s}{m_s} \frac{\tilde{\bm{\uprho}}_{s,i,j,k} + \tilde{\bm{\uprho}}_{s,i,j,k+1}}{2\Delta V}.
\end{gather}
Although the linear system from equation \eqref{eq:efieldeqn_disc_rhofieldcorr} is not obtained from a least action principle, it is a convenient approximation on account of its lower computational cost.

\subsection{Semi-discrete energy conservation}\label{sec:discreteenergycons}
For a periodic domain, the total magnetic field energy of the system can be defined as
\begin{equation}
    E_M = \frac 12 \tilde{\arr{H}} \cdot \arr{B}.
\end{equation}
Its rate of change becomes, using equation \eqref{eq:faraday_disc}
\begin{equation} \label{eq:MErate}
    \fract{E_M}{t} = \tilde{\arr{H}} \cdot \fract{\arr{B}}{t} = - \tilde{\arr{H}} \cdot \mathbb{C}\arrE.
\end{equation}
The total kinetic energy of the system is defined as
\begin{equation}
    E_K = \sum_{p=1}^{N_p} \left[\frac 12 w_p m_p \bv_p^2 \right].
\end{equation}
Using equations \eqref{eq:hodge_relns}, \eqref{eq:Vdotsdisc} and \eqref{eq:ampere_disc}, its rate of change becomes
\begin{align} \label{eq:KErate}
    \fract{E_K}{t} &= \sum_{p=1} \left[w_p q_p \bv_p \cdot \left(\bE^S(t,\bx_p) + \bv_p \times \bB^S(t,\bx_p)\right) \right] \nonumber
    = \sum_{p=1} w_p q_p \bv_p \cdot \bE^S(t,\bx_p) \notag \\
    & = \sum_{p=1} w_p q_p  \begin{pmatrix}
   v_{p,x} \\
   v_{p,y} \\
   v_{p,z}
   \end{pmatrix} \cdot
   \begin{pmatrix}
   \sum_{i,j,k} {\arr{E}}_{i+1/2,j,k} S_i^{c}(x_p) S_j^{n}(y_p) S_k^{n}(z_p) \\
   \sum_{i,j,k} {\arr{E}}_{i,j+1/2,k} S_i^{n}(x_p) S_j^{c}(y_p) S_k^{n}(z_p) \\
   \sum_{i,j,k} {\arr{E}}_{i,j,k+1/2} S_i^{n}(x_p) S_j^{n}(y_p) S_k^{c}(z_p)
   \end{pmatrix} \notag \\
    & = \sum_{i,j,k}\sum_{p=1} w_p q_p  \begin{pmatrix}
   v_{p,x} S_i^{c}(x_p) S_j^{n}(y_p) S_k^{n}(z_p) \\
   v_{p,y} S_i^{n}(x_p) S_j^{c}(y_p) S_k^{n}(z_p) \\
   v_{p,z} S_i^{n}(x_p) S_j^{n}(y_p) S_k^{c}(z_p)
   \end{pmatrix} \cdot
   \begin{pmatrix}
   {\arr{E}}_{i+1/2,j,k} \\
   {\arr{E}}_{i,j+1/2,k} \\
   {\arr{E}}_{i,j,k+1/2}
   \end{pmatrix} \notag \\
    & = \sum_{i,j,k} \begin{pmatrix}
   \tilde{\arr{J}}^x(t)_{i+1/2,j,k} \\
   \tilde{\arr{J}}^y(t)_{i,j+1/2,k} \\
   \tilde{\arr{J}}^z(t)_{i,j,k+1/2}
   \end{pmatrix} \cdot
   \begin{pmatrix}
   {\arr{E}}_{i+1/2,j,k} \\
   {\arr{E}}_{i,j+1/2,k} \\
   {\arr{E}}_{i,j,k+1/2}
   \end{pmatrix} \notag \\
   & = \tilde{\arr{J}} \cdot {\arr{E}}
   = (\mathbb{C}^\top\mathbb{H}_2\arrB) \cdot {\arr{E}} 
   = (\mathbb{C}^\top \tilde{\arr{H}}) \cdot {\arr{E}}
   = (\mathbb{C}^\top \tilde{\arr{H}})^\top {\arr{E}}
   = \tilde{\arr{H}}^\top \mathbb{C} {\arr{E}}
   = \tilde{\arr{H}} \cdot \mathbb{C} {\arr{E}}.
\end{align}
From equations \eqref{eq:MErate} and \eqref{eq:KErate}, it can be seen that the sum of magnetic and kinetic energies is conserved by our proposed semi-discrete formulation. It must be noted that the exact conservation of energy only holds true in the semi-discrete case. After discretizing in time, the energy conservation error would depend on the time-step size $\Delta t$, up to the order of accuracy of the discretization scheme. We also point out that the above condition of semi-discrete energy conservation exactly holds for periodic boundary conditions. This condition would not automatically hold true for non-periodic boundaries and would need modifications to account for boundary fluxes.
\section{Time-stepping scheme}\label{sec:timestepping}
We assume that particle positions and velocities are known at time $t_n$. This also includes the initial time $t = t_0 = 0$. To update the particle velocities and thus update their positions, we first need the electric and magnetic fields to calculate the forces acting on particles. The discrete Amp\`ere's equation \eqref{eq:ampere_disc} can be used to calculate the magnetic field. However, the discretized current density, $\tilde{\arr{J}}$, satisfied the divergence-free condition equation \eqref{eq:divJ_disc} only in the semi-discrete formulation that is continuous with respect to time. This divergence-free property does not hold for finitely sized time-steps. The divergence error would depend on the time-step size $\Delta t$, up to the order of accuracy of the scheme. Therefore, to proceed, $\tilde{\arr{J}}$ must first satisfy the divergence-free condition equation \eqref{eq:divJ_disc}. This is necessary for the quasineutrality assumption to hold true and also for obtaining a divergence-free magnetic field. This is achieved with the help of a `divergence-corrector' electric field, ${\bE}_\star$, and its corresponding potential, ${\psi}_\star$ where ${\bE}_\star = -\nabla {\psi}_\star$. We define their discrete counterparts on the primal grid edges and nodes and refer to them as ${\arr{E}}_\star$ and ${\arr{\uppsi}}_\star$, respectively. They also satisfy the condition
\begin{equation}\label{eq:defn_Estar}
    {\arr{E}}_\star = - \mathbb{G} {\arr{\uppsi}}_\star.
\end{equation}
The `corrector' electric field should be such that updating the particle velocities using the force exerted by this field over the time-step $\Delta t$ leads to a discretized current density $\tilde{\arr{J}}$ with zero discrete divergence, i.e. it would satisfy equation \eqref{eq:divJ_disc}. This gives
\begin{equation} \label{eq:divJ_Estar_disc}
\tilde{\mathbb{D}}\tilde{\arrJ} = \tilde{\mathbb{D}} \sum_p q_p w_p \tilde{\mathcal{R}}_2\left((\bv_{p,\star}(t) + \frac{q_p}{m_p} \bE_{\star}^S(t_n,\bx_p(t_n)) \Delta t)S(\bx-\bx_p(t))\right) = 0.
\end{equation}
Here, $\bE_{\star}^S(\bx_p) = (E_{\star,x}^S(\bx_p), E_{\star,y}^S(\bx_p), E_{\star,z}^S(\bx_p))$ is written in terms of the edge-integrals ${\arr{E}}_\star$ as
\begin{gather}
E_{\star,x}^S(\bx_p) = \sum_{i,j,k} {\arr{E}}_{\star,i+1/2,j,k} S_i^{c}(x_p) S_j^{n}(y_p) S_k^{n}(z_p), \\
E_{\star,y}^S(\bx_p) = \sum_{i,j,k} {\arr{E}}_{\star,i,j+1/2,k} S_i^{n}(x_p) S_j^{c}(y_p) S_k^{n}(z_p), \\
E_{\star,z}^S(\bx_p) = \sum_{i,j,k} {\arr{E}}_{\star,i,j,k+1/2} S_i^{n}(x_p) S_j^{n}(y_p) S_k^{c}(z_p).
\end{gather}
In terms of $\psi_\star$, \eqref{eq:divJ_Estar_disc} becomes
\begin{equation} \label{eq:div_psistar_disc}
\tilde{\mathbb{D}} \sum_p q_p w_p \tilde{\mathcal{R}}_2\left((\bv_{p,\star}(t) - \frac{q_p}{m_p} \nabla\psi_{\star}^S(t_n,\bx_p(t_n))\Delta t)S(\bx-\bx_p(t))\right) = 0.
\end{equation}
Defining an uncorrected current density $\tilde{\arr{J}}_\star$ in terms of the contributions from the uncorrected particle velocities $\bv_{p,\star}$, we get
\begin{equation}
\tilde{\arr{J}}_\star(t) = \sum_p q_p w_p \tilde{\mathcal{R}}_2\left( \bv_{p,\star}(t) S(\bx-\bx_p(t))\right).
\end{equation}
Substituting this in equation \eqref{eq:div_psistar_disc}, we get
\begin{equation} \label{eq:divJstar_psistar_disc}
\tilde{\mathbb{D}} \sum_p \frac{q_p^2}{m_p} w_p \tilde{\mathcal{R}}_2\left(\nabla\psi_{\star}^S(t_n,\bx_p(t_n))\Delta t S(\bx-\bx_p(t))\right) = \tilde{\mathbb{D}} \tilde{\arr{J}}_\star(t_n).
\end{equation}
The ${\arr{E}}_{\star}$ edge-integrals can be written in terms of ${\arr{\uppsi}}_\star$ as
\begin{gather}
E_{\star,x}^S(\bx_p) = -\nabla\psi_{\star,x}^S(\bx_p) = \sum_{i,j,k} ({\arr{\uppsi}}_{\star,i,j,k} - {\arr{\uppsi}}_{\star,i+1,j,k}) S_i^{c}(x_p) S_j^{n}(y_p) S_k^{n}(z_p), \\
E_{\star,y}^S(\bx_p) = -\nabla\psi_{\star,y}^S(\bx_p) = \sum_{i,j,k} ({\arr{\uppsi}}_{\star,i,j,k} - {\arr{\uppsi}}_{\star,i,j+1,k}) S_i^{n}(x_p) S_j^{c}(y_p) S_k^{n}(z_p), \\
E_{\star,z}^S(\bx_p) = -\nabla\psi_{\star,z}^S(\bx_p) = \sum_{i,j,k} ({\arr{\uppsi}}_{\star,i,j,k} - {\arr{\uppsi}}_{\star,i,j,k+1}) S_i^{n}(x_p) S_j^{n}(y_p) S_k^{c}(z_p).
\end{gather}
Hence, equation \eqref{eq:divJstar_psistar_disc} becomes a large, sparse linear system that is solved to obtain ${\arr{\uppsi}}_{\star}$, and ${\arrE}_{\star}$ is then obtained from equation \eqref{eq:defn_Estar}. The particle velocities are now updated as
\begin{equation} \label{eq:velocity_correction}
    \bv_p^n = \bv_{p,\star}^n + \frac{q_p}{m_p} \bE_\star^n(\bx_p) \Delta t.
\end{equation}
We now have the corrected velocities at time $t_n$ that give a divergence-free ${\arr{J}}$, calculated using equations \eqref{eq:depositjx}--\eqref{eq:depositjz}. This corrected ${\arr{J}}$ can now be used for calculating $\arrB$. The discretized vector potential $\arrA$ is now calculated using equation \eqref{eq:laplaceA_disc} and then $\arrB$ is calculated using the discretized magnetic field definition given in equation \eqref{eq:disc_field_defns} i.e. $\arrB = \mathbb{C}\arrA$. The electric field at time $t_n$ can now be calculated using the linear system in equation \eqref{eq:efieldeqn_disc}. All the large, sparse linear systems from equations \eqref{eq:laplaceA_disc}, \eqref{eq:efieldeqn_disc}, \eqref{eq:efieldeqn_disc_rhofieldcorr} and \eqref{eq:divJstar_psistar_disc} are solved in parallel using the \texttt{HYPRE} library \citep{falgout2002, falgout2006design, hypre}.

After the electric and magnetic fields at time $t_n$ are calculated, the positions and velocities of the particles of species $s$ are evolved using the following split scheme:

\begin{gather}
	\bv_p^- = \bv_p^n + \frac{q_p}{m_p} \bE^n(x_p^n) \frac{\Delta t}{2},\label{eq:vp_minus} \\
    \bv_p^{n+\frac{1}{2}} = \mathbb{R}(\bB^n(x_p^n),\; ((q_p/m_p)|\bB^n(x_p^n)|\Delta t/2)) \bv_p^-, \label{eq:vcrossB1}\\
    \bx_p^{n+1} = \bx_p^{n} + \bv_p^{n+\frac{1}{2}} \Delta t, \\
    \bv_p^{+} = \mathbb{R}(\bB^n(x_p^{n+1}),\;(q_p/m_p)|\bB^n(x_p^{n+1})|\Delta t/2) \bv_p^{n+\frac{1}{2}}, \label{eq:vcrossB2}\\
	\bv_p^{n+1} = \bv_p^{+} + \frac{q_p}{m_p} \bE^n(x_p^{n+1}) \frac{\Delta t}{2} \label{eq:vp_plus}.
\end{gather}
Equations \eqref{eq:vcrossB1} and \eqref{eq:vcrossB2} are $\bv \times \bB$ rotation updates, where the magnitude of $\bv$ is retained after the update. The rotation matrix $\mathbb{R}(\bB, \alpha)$ where $\bB$ is the magnetic field vector and $\alpha$ is a dimensionless parameter, is given by
\begin{equation}\label{eq:rotn_matrix}
	\mathbb{R}(\bB, \alpha) = \begin{bmatrix}
   b_x b_x + (b_y b_y + b_z b_z)c_\alpha & b_z s_\alpha + b_x b_y (1-c_\alpha) & -b_y s_\alpha + b_x b_z (1-c_\alpha) \\
   -b_z s_\alpha + b_y b_x (1-c_\alpha) & b_y b_y + (b_x b_x + b_z b_z)c_\alpha & b_x s_\alpha + b_y b_z (1-c_\alpha) \\
   b_y s_\alpha + b_x b_z (1-c_\alpha) & -b_x s_\alpha + b_y b_z (1-c_\alpha) & b_z b_z + (b_x b_x + b_y b_y)c_\alpha
   \end{bmatrix}.
\end{equation}
Here, $c_\alpha = \cos(\alpha)$, $s_\alpha = \sin(\alpha)$ and ${\bf b} = (b_x, b_y, b_z) = \bB/|\bB|$, i.e. the unit vector in the direction of the magnetic field. Once the particle positions and velocities are available at time $t_{n+1}$, the entire procedure described above is repeated for the next time-step. This time-stepping scheme is second order accurate. For better readability, the time-stepping scheme is summarized as follows:
\begin{enumerate}
    \item Particle positions and velocities are available at time $t_n$. Equation \eqref{eq:divJstar_psistar_disc} is solved to obtain The `divergence-corrector' potential, ${\arr{\uppsi}}_{\star}$. The `divergence-corrector' electric field, ${\arrE}_{\star}$, is then obtained by solving equation \eqref{eq:defn_Estar}.
    \item Particle velocities are updated using equation \eqref{eq:velocity_correction}.
    \item The divergence-free discrete current density, $\arr{J}$, is calculated using equations \eqref{eq:depositjx}--\eqref{eq:depositjz}.
    \item The discretized vector potential $\arrA$ is calculated from equation \eqref{eq:laplaceA_disc} and $\arrB$ is calculated using equation \eqref{eq:disc_field_defns}.
    \item $\arr{E}$ is calculated using equation \eqref{eq:efieldeqn_disc}.
    \item Particle positions and velocities at time $t_{n+1}$ are calculated using the standard second order Boris-push as per equations \eqref{eq:vp_minus}--\eqref{eq:vp_plus}, completing the time-step.
\end{enumerate}
The time-step restriction for the time-stepping scheme is given by
\begin{equation}
\Delta t \leq \frac{C}{|\Omega_{c,e}|},
\end{equation}
where $C$ is a constant. This is because the electron cyclotron wave with frequency $\Omega_{c,e}$ is the fastest wave in the quasineutral model.

%%%%%%%%%%%%%%%%%%%%%%%%%%%%%%%%%%%%%%%%%%%%%%%%%%%%%%%%%
\section{Dispersion relation}\label{sec:dispersionrelation}
It is possible to perform a linear perturbation analysis of the Vlasov-Maxwell system to obtain a dispersion relation describing the various eigenmodes generated by the system. This has been shown by various authors such as \citet{brambilla1998kinetic}, \citet{fitzpatrick2022plasma} and \citet{stix1992waves}, to name a few. It must be noted that while these authors derive the dispersion relation for the non-quasineutral, full Vlasov-Maxwell system, the dispersion relation required here can be obtained from the full system by simply taking the quasineutral limit i.e. $\epsilon_0 \to 0$. However, for the sake of completeness, we still describe the steps required to obtain the dispersion relation for the quasineutral Vlasov-Maxwell system. An equilibrium state is assumed where $\bE = \bE_0$, $\bB = \bB_{ext}$ and $f_s = f_{s,0}$. Perturbations from equilibrium quantities are denoted by the subscript `1'. To obtain the dispersion relation, plane-wave perturbations of the form
\begin{equation}\label{eq:perturbation}
\bE_1 = \hat{\bE} e^{i(\bk \cdot \bx-\omega t)}
\end{equation}
can be assumed, where $\bk$ is the wavenumber and $\omega$ is the angular frequency for the plane-wave perturbation. Applying such perturbations to \eqref{eq:J}, \eqref{eq:qnVlasov}, \eqref{eq:qnampere} and \eqref{eq:qnfaraday}, we get
\begin{gather}
    \hat{\bJ} = \sum_{s} q_s \int \hat{f_s} \bv d\bv, \label{eq:J_perp} \\
    -i \omega \hat{f_s} + i\fract{\bX}{t}\cdot \bk \hat{f_s} + \frac{q_s}{m_s}\left[{(\bV \times \bB_{ext})} \cdot\fracp{\hat{f_s}}{\bV} + {(\hat{\bE} + \bV \times \hat{\bB})} \cdot\fracp{f_{s,0}}{\bV} \right] =0, \label{eq:qnVlasov_perp} \\
    i \bk \times \hat{\bB} = \mu_0 \hat{\bJ}, \label{eq:qnampere_perp}\\
    \omega\hat{\bB} = \bk \times \hat{\bE}. \label{eq:qnfaraday_perp}
\end{gather}
Eliminating $\hat{\bB}$, $\hat{f_s}$ and $\hat{\bJ}$ from these equations, we obtain the equation
\begin{equation} \label{eq:Ehat1}
     \bk \times \bk \times \hat{\bE} + (\omega/c)^2 \underline{\underline{\boldsymbol{\epsilon}}} \cdot \hat{\bE} = 0,
\end{equation}
where $\underline{\underline{\epsilon}}$ is the dielectric tensor of the plasma, that can be written as
\begin{equation}\label{eq:epsilon}
	\underline{\underline{\boldsymbol{\epsilon}}} = \begin{bmatrix}
   \epsilon_{xx} & \epsilon_{xy} & \epsilon_{xz} \\
   \epsilon_{yx} & \epsilon_{yy} & \epsilon_{yz} \\
   \epsilon_{zx} & \epsilon_{zy} & \epsilon_{zz}
   \end{bmatrix}.
\end{equation}
The plasma dielectric tensor is a matrix that describes the behaviour of a magnetized plasma in the presence of an electric field. It determines the dispersion relation and contains all the information on linear wave propagation in a uniform magnetized plasma. Denoting the magnitude of $\bk$ as $k$, we can define the refractive index $n = kc/\omega$ and the wavevector $\boldsymbol{\kappa} = \bk/k$, equation \eqref{eq:Ehat1} becomes
\begin{equation} \label{eq:Ehat2}
     n^2 \boldsymbol{\kappa} \times \boldsymbol{\kappa} \times \hat{\bE} + \underline{\underline{\boldsymbol{\epsilon}}} \cdot \hat{\bE} = 0.
\end{equation}
Taking the determinant of this linear system gives us the hot plasma dispersion relation
\begin{equation}\label{eq:dispreln}
\det|n^2(\kappa_i \kappa_j - \delta_{ij}) + \epsilon_{ij}| = 0.
\end{equation}
Given the equivalence of all directions perpendicular to the static background magnetic field, we write $\underline{\underline{\boldsymbol{\epsilon}}}$ in terms of $k_{\perp}$ and $k_{\parallel}$, which are the components of $\bk$ perpendicular and parallel to the magnetic field, respectively. We assume $\theta$ to be the angle between $\bk$ and the static background magnetic field. Without loss of generality, we define $k_x = k_{\perp}$, $k_y = 0$ and $k_z = k_{\parallel}$, and therefore get
\begin{gather}
    k_x = k_{\perp} = k \sin(\theta) \Longrightarrow \kappa_x = \sin(\theta), \label{eq:k_perp} \\
    k_z = k_{\parallel} = k \cos(\theta) \Longrightarrow \kappa_z = \cos(\theta). \label{eq:k_par}
\end{gather}
The background magnetic field is therefore along the $z$-direction. The hot plasma dispersion relation, thus becomes
\begin{equation}\label{eq:hot_dispreln}
	\det\begin{bmatrix}
   \epsilon_{xx} - n^2\cos^2(\theta) & \epsilon_{xy} & \epsilon_{xz} + n^2\cos(\theta)\sin(\theta)\\
   \epsilon_{yx} & \epsilon_{yy} - n^2 & \epsilon_{yz} \\
   \epsilon_{zx} + n^2\cos(\theta)\sin(\theta) & \epsilon_{zy} & \epsilon_{zz} - n^2\sin^2(\theta)
   \end{bmatrix} = 0.
\end{equation}
We also define $n_{\perp} = n\sin(\theta)$ and $n_{\parallel} = n\cos(\theta)$.
In the quasineutral case, the individual terms of the dielectric tensor $\underline{\underline{\boldsymbol{\epsilon}}}$ are given by

\begin{gather}
\epsilon_{xx} = - \sum_{s} \frac{\omega_{p,s}^2}{\omega^2} \sum_{n=-\infty}^{n=+\infty} \frac{n^2}{\lambda_s} I_n(\lambda_s) e^{-\lambda_s}(-x_{0,s}Z(x_{n,s})), \label{eq:epsilonxx} \\
\epsilon_{xy} = -\epsilon_{yx} = -i \sum_{s} \frac{\omega_{p,s}^2}{\omega^2} \sum_{n=-\infty}^{n=+\infty} n[I'_n(\lambda_s) - I_n(\lambda_s)] e^{-\lambda_s}(-x_{0,s}Z(x_{n,s})), \label{eq:epsilonxy} \\
\epsilon_{xz} = \epsilon_{zx} = -\half n_{\perp} n_{\parallel} \sum_{s} \frac{\omega_{p,s}^2}{\omega \Omega_{c,s}} \frac{v_{th,s}^2}{c^2} \sum_{n=-\infty}^{n=+\infty} \frac{n}{\lambda_s} I_n(\lambda_s) e^{-\lambda_s} (x_{0,s}^2 Z'(x_{n,s})), \label{eq:epsilonxz} \\
\epsilon_{yy} = - \sum_{s} \frac{\omega_{p,s}^2}{\omega^2} \sum_{n=-\infty}^{n=+\infty} \left[ \frac{n^2}{\lambda_s} I_n(\lambda_s) - 2\lambda_s [I'_n(\lambda_s) - I_n(\lambda_s)] \right] e^{-\lambda_s}(-x_{0,s}Z(x_{n,s})), \label{eq:epsilonyy} \\
\epsilon_{yz} = -\epsilon_{zy} = \frac{i}{2} n_{\perp} n_{\parallel} \sum_{s} \frac{\omega_{p,s}^2}{\omega \Omega_{c,s}} \frac{v_{th,s}^2}{c^2} \sum_{n=-\infty}^{n=+\infty} [I'_n(\lambda_s) - I_n(\lambda_s)] e^{-\lambda_s} (x_{0,s}^2 Z'(x_{n,s})), \label{eq:epsilonyz} \\
\epsilon_{zz} = - \sum_{s} \frac{\omega_{p,s}^2}{\omega^2} \sum_{n=-\infty}^{n=+\infty} I_n(\lambda_s)e^{-\lambda_s}(x_{0,s} x_{n,s} Z'(x_{n,s})). \label{eq:epsilonzz}
\end{gather}
In the above expressions, $\lambda_s$ and $x_{n,s}$ are dimensionless quantities given by
\begin{equation}\label{eq:lambda}
    \lambda_s = \frac{k_{\perp}^2 v_{th,s}^2}{2 \Omega_{c,s}^2},
\end{equation}
and
\begin{equation}\label{eq:xns}
x_{n,s} = \frac{\omega - n\Omega_{c,s}}{k_{\parallel} v_{th,s}}.
\end{equation}
\nmn{The terms $\omega_{p,s}$ and $\Omega_{c,s}$ are the plasma and cyclotron frequencies of species $s$. These frequencies, for the two species, are given by
\begin{equation}\label{eq:omegas}
\omega_{p,i} = \sqrt{\frac{Z_i^2 n_i e^2}{\epsilon_0 m_i}}, ~~~~ \omega_{p,e} = \sqrt{\frac{n_e e^2}{\epsilon_0 m_e}}
\end{equation}
and
\begin{equation}\label{eq:Omegas}
\Omega_{c,i} = \frac{e Z_i B_{ext}}{m_i}, ~~~~ \Omega_{c,e} = -\frac{e B_{ext}}{m_e},
\end{equation}
where $Z_i$ is the ion atomic number, $e$ is the electronic charge, and $B_{ext}$ is the magnetic field magnitude. Note that the electron cyclotron frequency, $\Omega_{c,e}$, is defined as a negative value. The terms $n_i$ and $n_e$ are the ion and electron particle number density, respectively. Similarly, $m_i$ and $m_i$ denote the respective particle masses. Also, $v_{th,s}$ is the thermal velocity of the species $s$ given by
\begin{equation}\label{eq:vths}
v_{th,s} = \sqrt{\frac{2T_s}{m_s}},
\end{equation}
where $T_s$ is the temperature of the species.
Also, $Z$ is the plasma dispersion function and $I_n$ denotes the modified Bessel function of order $n$.}

\subsection{Cold plasma approximation}\label{sec:dispersionrelationcold}
To obtain a simplified dispersion relation for the case of a cold plasma, we assume the limit $T_s \rightarrow 0$. This also implies $\lambda_s \rightarrow 0$ and $|x_{n,s}| \rightarrow \infty$, causing the components $\epsilon_{xz}$, $\epsilon_{zx}$, $\epsilon_{yz}$ and $\epsilon_{zy}$ of the dielectric tensor to vanish, i.e.
\begin{equation} \label{eq:coldplasma_zeros}
\lim_{T_s \rightarrow 0} \epsilon_{xz} = \lim_{T_s \rightarrow 0} \epsilon_{zx} = \lim_{T_s \rightarrow 0} \epsilon_{yz} = \lim_{T_s \rightarrow 0} \epsilon_{zy} = 0.
\end{equation}
The remaining terms simplify as
\begin{gather}
\lim_{T_s \rightarrow 0} \epsilon_{xx} = \lim_{T_s \rightarrow 0} \epsilon_{yy} = S = \half (R+L) = - \sum_{s} \frac{\omega_{p,s}^2}{\omega^2 - \Omega_{c,s}^2}, \label{eq:coldplasma_epsilonxx} \\
\lim_{T_s \rightarrow 0} \epsilon_{xy} = -\lim_{T_s \rightarrow 0} \epsilon_{yx} = -iD = \frac{1}{2i} (R-L) = - i \sum_{s} \frac{\Omega_{c,s}}{\omega} \frac{\omega_{p,s}^2}{\omega^2 - \Omega_{c,s}^2}, \label{eq:coldplasma_epsilonxy} \\
\lim_{T_s \rightarrow 0} \epsilon_{zz} = P = - \sum_{s} \frac{\omega_{p,s}^2}{\omega^2}. \label{eq:coldplasma_epsilonzz}
\end{gather}
Here the terms $R$ and $L$ are given by
\begin{gather}
R = - \sum_{s} \frac{\omega_{p,s}^2}{\omega(\omega + \Omega_{c,s})}, \label{eq:coldplasma_R} \\
L = - \sum_{s} \frac{\omega_{p,s}^2}{\omega(\omega - \Omega_{c,s})}. \label{eq:coldplasma_L}
\end{gather}
Therefore, for a cold plasma, equation \eqref{eq:hot_dispreln} simplifies to
\begin{equation}\label{eq:cold_dispreln}
	\det\begin{bmatrix}
   S - n^2\cos^2(\theta) & -iD & n^2\cos(\theta)\sin(\theta)\\
   iD & S - n^2 & 0 \\
   n^2\cos(\theta)\sin(\theta) & 0 & P - n^2\sin^2(\theta)
   \end{bmatrix} = 0.
\end{equation}
This is the cold plasma dispersion relation. This approximation is helpful also in the case of warm plasmas, in obtaining simplified expressions for various wave frequencies. In the following two subsections, we analyze the warm and cold plasma dispersion relations to describe the waves that are obtained propagating in directions parallel and perpendicular to the background equilibrium magnetic field. Wherever possible, we use the cold plasma dispersion relation to obtain simplified expressions for the $k$-$\omega$ relations. This works well for waves existing in cold plasmas. In other cases, i.e. for waves existing only in warm plasmas, we use the warm plasma dispersion relation to describe them.

\subsection{Waves $\parallel$ to $\bB_{ext}$}\label{sec:parallelwaves}
The dispersion relation for waves propagating in a direction parallel the background magnetic field can be obtained by substituting $\theta = 0$ in equation \eqref{eq:cold_dispreln}. We thus get the equation
\begin{equation}\label{eq:cold_dispreln_par}
	\det\begin{bmatrix}
   S - n^2 & -iD & 0\\
   iD & S - n^2 & 0 \\
   0 & 0 & P
   \end{bmatrix} = 0.
\end{equation}
For the quasineutral case, there is no solution to the equation $P = 0$. For the non-quasineutral case, the solution for this equation would have been the Langmuir wave. The remaining factor simplifies to the equations
\begin{gather}
n^2 = S + D = R, \label{eq:coldplasma_par_R} \\
n^2 = S - D = L. \label{eq:coldplasma_par_L}
\end{gather}
For the R-mode wave in equation \eqref{eq:coldplasma_par_R}, since $\Omega_{c,e}$ is negative, a positive solution for $\omega$ exists only for $\omega < |\Omega_{c,e}|$. It is clear that $n \rightarrow +\infty$ as $\omega \rightarrow |\Omega_{c,e}|^-$. This is a right-handed, transverse, circularly polarized wave, called the electron cyclotron wave (ECW), with the eigenvector $(E_x, iE_x, 0)$. Similarly, for the L-mode in equation \eqref{eq:coldplasma_par_L}, a positive solution for $\omega$ exists only for $\omega < |\Omega_{c,i}|$. In this case, $n \rightarrow +\infty$ as $\omega \rightarrow |\Omega_{c,i}|^-$. This is a left-handed, transverse, circularly polarized wave and the corresponding eigenvector is $(E_x, -iE_x, 0)$. This is called the ion cyclotron wave (ICW). Both these waves are seen below in Figures \ref{fig:kpar_double_Eperp_Ez} and \ref{fig:kpar_double_Eperp_Ey} in Section \ref{sec:sim_parallelwaves_double}. For a single-species plasma with kinetic electrons and stationary ions in the background, only the electron cyclotron wave (ECW) is observed, as can be seen in Figures \ref{fig:kpar_single_Eperp_Ez} and \ref{fig:kpar_single_Eperp_Ey} in Section \ref{sec:sim_parallelwaves_single}. The full, non-quasineutral Vlasov-Maxwell system generates two more solutions with components $E_x$ and $E_y$, from the system of equations \eqref{eq:coldplasma_par_R} and \eqref{eq:coldplasma_par_L}. These are a left-handed and a right-handed circularly polarized electromagnetic wave wherein $\omega \to ck$ as $k \to \infty$.

The above cyclotron waves have been obtained from the cold plasma dispersion relation. However, there are other modes that do not exist in the cold plasma limit. Describing these waves requires the more generic hot plasma dispersion relation given in equation \eqref{eq:hot_dispreln}. Substituting $k_{\perp} = 0$ in the plasma dielectric tensor, the terms $\epsilon_{xz}$, $\epsilon_{yz}$, $\epsilon_{zx}$ and $\epsilon_{zy}$ simplify to 0, while the other terms are given by
\begin{gather}
\epsilon_{xx} = \epsilon_{yy} = \half \sum_{s} \frac{\omega_{p,s}^2}{\omega k_{\parallel} v_{th,s}} \left[Z\left(\frac{\omega - \Omega_{c,s}}{k_{\parallel} v_{th,s}}\right) + Z\left(\frac{\omega + \Omega_{c,s}}{k_{\parallel} v_{th,s}}\right) \right], \label{eq:epsilonxx_par} \\
\epsilon_{xy} = -\epsilon_{yx} = \frac{i}{2}\sum_{s} \frac{\omega_{p,s}^2}{\omega k_{\parallel} v_{th,s}} \left[Z\left(\frac{\omega - \Omega_{c,s}}{k_{\parallel} v_{th,s}}\right) - Z\left(\frac{\omega + \Omega_{c,s}}{k_{\parallel} v_{th,s}}\right) \right], \label{eq:epsilonxy_par} \\
\epsilon_{zz} = - \sum_{s} \frac{\omega_{p,s}^2}{(k_{\parallel} v_{th,s})^2} Z'\left(\frac{\omega}{k_{\parallel} v_{th,s}}\right). \label{eq:epsilonzz_par}
\end{gather}
The dispersion relation becomes
\begin{equation}\label{eq:hot_dispreln_par}
	((\epsilon_{xx}-n^2)(\epsilon_{yy}-n^2) - \epsilon_{xy} \epsilon_{yx})\epsilon_{zz} = 0.
\end{equation}
Besides the warm plasma generalizations of the cyclotron waves described above, the other solutions of this equation are the heavily damped, higher-order modes of Alfv\'en-cyclotron waves that have been studied by various researchers such as \citet{araneda2012interactions}, \citet{astudillo1996high} and \citet{Matsuda_1986}. The modes obtained from the factor $((\epsilon_{xx}-n^2)(\epsilon_{yy}-n^2) - \epsilon_{xy} \epsilon_{yx}) = 0$, with eigenvector $(E_x, E_y, 0)$ manifest as straight lines emanating from ($k$, $\omega$) = (0, $\Omega_{c,e}$), forming a cone-shaped structure on the $k$-$\omega$ plot. For a double-species plasma with ions, these modes also exist for the second cyclotron frequency, thus emanating from ($k$, $\omega$) = (0, $\Omega_{c,i}$). This structure can be clearly seen in Figures \ref{fig:kpar_single_Eperp_Ez} and \ref{fig:kpar_single_Eperp_Ey} in Section \ref{sec:sim_parallelwaves_single} and Figures \ref{fig:kpar_double_Eperp_Ez} and \ref{fig:kpar_double_Eperp_Ey} in Section \ref{sec:sim_parallelwaves_double}. The mode obtained from $\epsilon_{zz} = 0$, with eigenvector $(0,0,E_z)$ also forms a similar cone-shaped structure emanating from ($k$, $\omega$) = (0, 0). This can be seen in Figures \ref{fig:kpar_single_Epar} and \ref{fig:kpar_double_Epar}. These higher-order modes are a thermal effect wherein the cone-angle decreases with temperature, reducing to zero in the cold plasma limit.

\subsection{Waves $\perp$ to $\bB_{ext}$}\label{sec:perpendicularwaves}
The dispersion relation for waves propagating in a direction parallel the background magnetic field can be obtained by substituting $\theta = \pi/2$ in equation \eqref{eq:cold_dispreln}. We thus get the equation
\begin{equation}\label{eq:cold_dispreln_perp}
	\det\begin{bmatrix}
   S & -iD & 0\\
   iD & S - n^2 & 0 \\
   0 & 0 & P - n^2
   \end{bmatrix} = 0.
\end{equation}
Again, for the quasineutral case, there is no solution for $P-n^2 = 0$. For the non-quasineutral case, this would have resulted in what is known as the ordinary or O-mode, which is an electromagnetic plasma wave with the component $E_z$. The remaining factor simplifies to
\begin{equation}\label{eq:cold_plasma_perp}
	n^2 = \frac{RL}{S}.
\end{equation}
For a single-species plasma with kinetic electrons and stationary ions in the background, this reduces to
\begin{equation}\label{eq:cold_plasma_perp_single}
	n^2 = -\frac{\omega_{p,e}^2}{\omega^2},
\end{equation}
which has no solution. For the two-species case with kinetic electrons as well as ions, there exists a solution for $\omega < \Omega_H$, where $\Omega_H$ is termed as the hybrid frequency and is given by
\begin{equation}\label{eq:hybrid_frequency}
	\Omega_H = \sqrt{\frac{\Omega_{c,i}^2\omega_{p,e}^2 + \Omega_{c,e}^2\omega_{p,i}^2}{\omega_{p,e}^2 + \omega_{p,i}^2}}.
\end{equation}
This value can be found by limiting equation \eqref{eq:cold_plasma_perp} to $\infty$ i.e. $S = 0$. It can also be seen that $n \rightarrow +\infty$ as $\omega \rightarrow |\Omega_H|^-$. The eigenvector for this wave is $(E_x, -i(S/D)E_x, 0)$. We see this wave in Figures \ref{fig:kperp_double_Eperp_Ex}, \ref{fig:kperp_double_Eperp_hot_Ex}, \ref{fig:kperp_double_Eperp_Ey}, \ref{fig:kperp_double_Eperp_hot_Ey}, \ref{fig:kperp_double_Eperp_Ey_mag} and \ref{fig:kperp_double_Eperp_hot_Ey_mag} in Section \ref{sec:sim_perpendicularwaves_double} below. For the full, non-quasineutral Vlasov-Maxwell system, equation \eqref{eq:cold_plasma_perp} produces three more solutions. For a two-species plasma, there is a second hybrid wave, wherein the eigenfrequency $\omega$ approaches a constant value as $k \to \infty$. There are also two more solutions that are called the extraordinary or X-modes. These X-modes are electromagnetic modes wherein $\omega \to ck$ as $k \to \infty$.

We now use the hot plasma dispersion relation to look at waves that do not exist in the cold plasma limit. Substituting $k_{\parallel} = 0$ in the plasma dielectric tensor, the terms $\epsilon_{xz}$, $\epsilon_{yz}$, $\epsilon_{zx}$ and $\epsilon_{zy}$ simplify to 0, while the other terms are given by
\begin{gather}
\epsilon_{xx} = - \sum_{s} \frac{\omega_{p,s}^2}{\omega} \frac{e^{-\lambda_s}}{\lambda_s} \sum_{n=-\infty}^{n=+\infty} \frac{n^2 I_n(\lambda_s)}{\omega - n \Omega_{c,s}}, \label{eq:epsilonxx_perp} \\
\epsilon_{xy} = -\epsilon_{yx} = -i \sum_{s} \frac{\omega_{p,s}^2}{\omega} e^{-\lambda_s}  \sum_{n=-\infty}^{n=+\infty} \frac{n[I'_n(\lambda_s) - I_n(\lambda_s)]}{\omega - n \Omega_{c,s}}, \label{eq:epsilonxy_perp} \\
\epsilon_{yy} = - \sum_{s} \frac{\omega_{p,s}^2}{\omega} \frac{e^{-\lambda_s}}{\lambda_s} \sum_{n=-\infty}^{n=+\infty} \frac{n^2 I_n(\lambda_s) + 2\lambda_s^2 I_n(\lambda_s) - 2\lambda_s^2 I'_n(\lambda_s)}{\omega - n \Omega_{c,s}}, \label{eq:epsilonyy_perp} \\
\epsilon_{zz} = -\sum_{s} \frac{\omega_{p,s}^2}{\omega} e^{-\lambda_s}  \sum_{n=-\infty}^{n=+\infty} \frac{I_n(\lambda_s)}{\omega - n \Omega_{c,s}}. \label{eq:epsilonzz_perp}
\end{gather}
The dispersion relation becomes
\begin{equation}\label{eq:hot_dispreln_perp}
	(\epsilon_{xx} (\epsilon_{yy}-n^2) - \epsilon_{xy} \epsilon_{yx})(\epsilon_{zz}-n^2) = 0.
\end{equation}
We first consider the equation $(\epsilon_{zz}-n^2) = 0$, with eigenvector $(0, 0, E_z)$. From equation \eqref{eq:epsilonzz_perp}, it can be clearly seen that the refractive index $n = n_{\perp}$ becomes infinite at all harmonics of the cyclotron frequencies i.e. $\omega = m  \Omega_{c,s}$
where $m$ is a positive integer and $s$ can be electrons or ions. Besides these cyclotron harmonic resonances, there are no wave solutions to this equation. Without the quasineutral assumption, an electromagnetic plasma wave would have been a solution to this equation. The other factor, i.e. $(\epsilon_{xx} (\epsilon_{yy}-n^2) - \epsilon_{xy} \epsilon_{yx}) = 0$, also gives rise to these resonances on account of the $(\omega - n\Omega_{c,s})$ terms in the denominators seen on the RHS of equations \eqref{eq:epsilonxx_perp}--\eqref{eq:epsilonyy_perp}. The other solutions to this equation are the Bernstein waves \citep{bernstein} with the associated eigenvector $(E_x, E_y, 0)$. We note that the $x$-direction is the direction of wave propagation, and therefore the $E_x$ Bernstein waves are longitudinal in nature, while the $E_y$ Bernstein waves are transverse. The cyclotron harmonic resonances can be seen in Figures \eqref{fig:kperp_single} and \eqref{fig:kperp_double}
 in the wave spectra of all field components. The Bernstein waves can be clearly seen in Figures \eqref{fig:kperp_single_Eperp_hot_Ex}, \eqref{fig:kperp_single_Eperp_hot_Ey}, \eqref{fig:kperp_double_Eperp_hot_Ex} and \eqref{fig:kperp_double_Eperp_hot_Ey}. The Bernstein waves also exist in Figures \eqref{fig:kperp_single_Eperp_Ex}, \eqref{fig:kperp_single_Eperp_Ey}, \eqref{fig:kperp_double_Eperp_Ex} and \eqref{fig:kperp_double_Eperp_Ey} but cannot be distinctly seen on account of overlap with the cyclotron harmonic resonances.
%%%%%%%%%%%%%%%%%%%%%%%%%%%%%%%%%%%%%%%%%%%%%%%%%%%%%%%%%

%%%%%%%%%%%%%%%%%%%%%%%%%%%%%%%%%%%%%%%%%%%%%%%%%%%%%%%%%
\section{Numerical tests}\label{sec:tests}
In this section, we perform various tests to validate our algorithm, by drawing comparisons between the various waves obtained from simulations and the corresponding analytical relations obtained from the dispersion relation. In these tests, a uniform quasineutral plasma with Gaussian distributions for each species, under constant initial background magnetic fields is considered. The simulation is allowed to run for a long time and the Fast-Fourier Transform (FFT) of the resultant electric field time-series data is obtained. The FFT spectrum can then be plotted on a $k$-$\omega$ plot along with their analytical counterparts. For all these simulations, the particles are generated using a quasi-random Sobol sampler. The computational domain is a quasi-1D domain with domain size $[0,10]\times[0,1]\times[0,1]$ and a computational grid of $256\times8\times8$ cells. The higher 256-cell resolution is taken in the direction along which the wave spectrum is to be studied. Periodic boundary conditions are applied at all boundaries. Next, we perform a test where a single sinusoidal perturbation is initialized and its damping rate is compared with the theoretical result obtained from the dispersion relation. Finally, we conduct studies to assess the scalability of the code.

\subsection{Waves propagating $\parallel$ to $\bB_{ext}$}\label{sec:sim_parallelwaves}

\subsubsection{Single-species plasma}\label{sec:sim_parallelwaves_single}
\begin{figure}
    \centering
    % First row
    \begin{subfigure}[b]{0.67\textwidth}
        \centering
        \includegraphics[width=\linewidth, trim=15pt 0pt 20pt 0pt, clip]{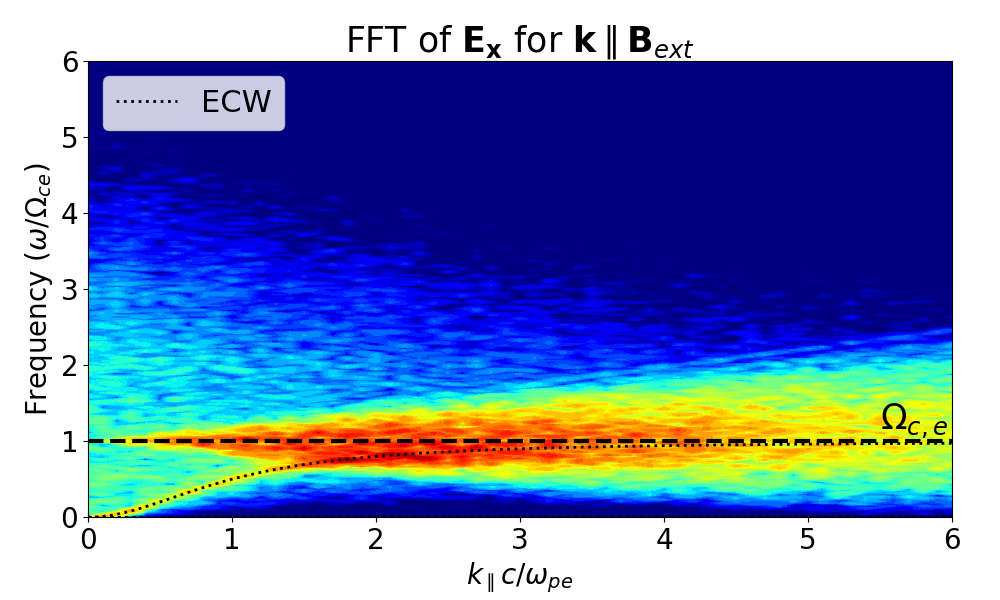}
        \caption{}
        \label{fig:kpar_single_Eperp_Ey}
    \end{subfigure}
    \hfill
    \begin{subfigure}[b]{0.67\textwidth}
        \centering
        \includegraphics[width=\linewidth, trim=15pt 0pt 20pt 0pt, clip]{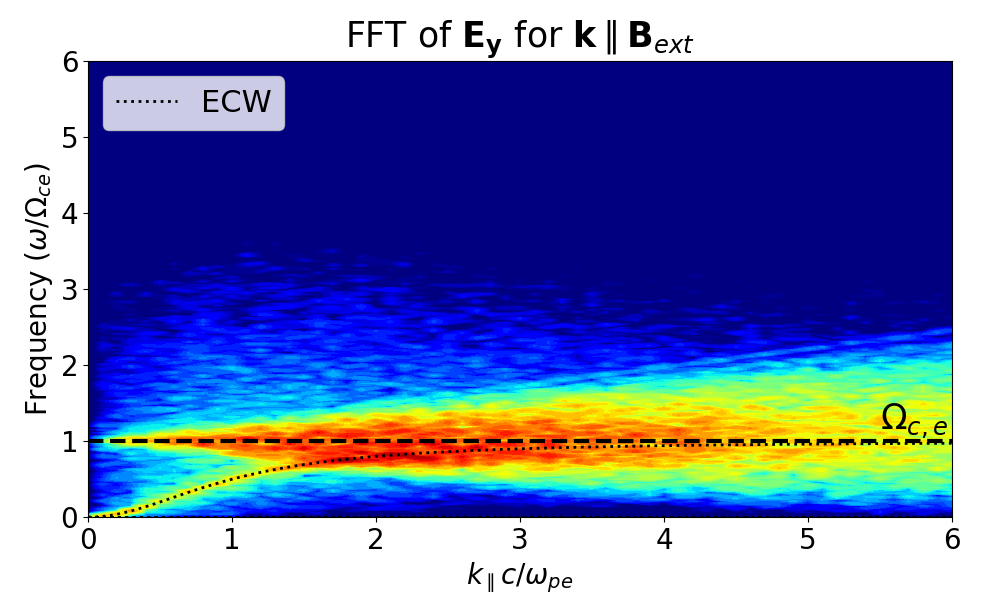}
        \caption{}
        \label{fig:kpar_single_Eperp_Ez}
    \end{subfigure}
    \hfill
    \begin{subfigure}[b]{0.67\textwidth}
        \centering
        \includegraphics[width=\linewidth, trim=15pt 0pt 20pt 0pt, clip]{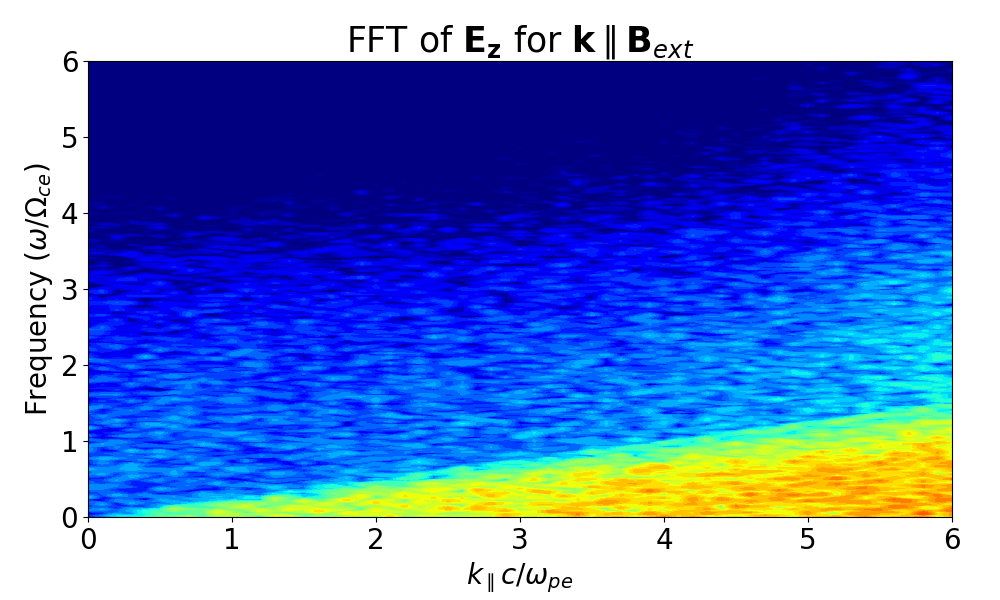}
        \caption{}
        \label{fig:kpar_single_Epar}
    \end{subfigure}
    %\vspace{8pt}
    \caption{Wave spectra of (a) $E_x$, (b) $E_y$ and (c) $E_z(E_\parallel)$ for a single-species electron-only plasma with $v_{th,e} = 0.05c$, for $\bk \parallel \bB_{ext}$. $E_x$ and $E_y$ waves are transverse and $E_z(E_\parallel)$ waves are longitudinal. In (a) and (b), the dotted line shows the electron cyclotron wave (ECW) while the thick dashed line shows the electron cyclotron frequency $\Omega_{c,e}$.}
    \label{fig:kpar_single_Eperp}
\end{figure}

A single-species simulation with kinetic electrons and stationary ions in the background is first considered. A total of 1000 particles per cell for the electron species is used. The thermal velocity of the electrons is taken to be $v_{th,e} = 0.05c$. As we are studying the spectrum of waves parallel to the direction of the background magnetic field, the 256-cell resolution is along this direction. The normalized background magnetic field is $B_{ext} = 1$. The various waves are obtained as a result of the random noise generated from the particle distribution. The wave spectra for the $E_x$ and $E_y$ fields are shown in Figures \ref{fig:kpar_single_Eperp_Ey} and \ref{fig:kpar_single_Eperp_Ez}, respectively. The electron cyclotron wave (ECW) is clearly observed here, with the analytical result from equation \eqref{eq:coldplasma_par_R} and the electron cyclotron frequency $\Omega_{c,e}$ superimposed on it. The $E_z(E_\parallel)$ wave spectrum is shown in Figure \eqref{fig:kpar_single_Epar}. The conical structure of the damped higher-order Alfv\'en-cyclotron modes described in Section \ref{sec:parallelwaves} is also clearly visible in the wave spectra of all three field components. It must be noted that the results for the wave spectrum of the perpendicular component of the electric field is almost identical in both the perpendicular directions. On account of the quasineutrality assumption, the Langmuir wave is missing in the $E_z(E_\parallel)$ spectrum in Figure \eqref{fig:kpar_single_Epar} and the electromagnetic L-mode and R-mode are missing in Figures \ref{fig:kpar_single_Eperp_Ey} and \ref{fig:kpar_single_Eperp_Ez}. These missing modes can be observed clearly in the works by \citet{meng2025} and \citet{kormann2024}.

\subsubsection{Double-species plasma}\label{sec:sim_parallelwaves_double}
\begin{figure}
    \centering
    % First row
    \begin{subfigure}[b]{0.67\textwidth}
        \centering
        \includegraphics[width=\linewidth, trim=15pt 0pt 20pt 0pt, clip]{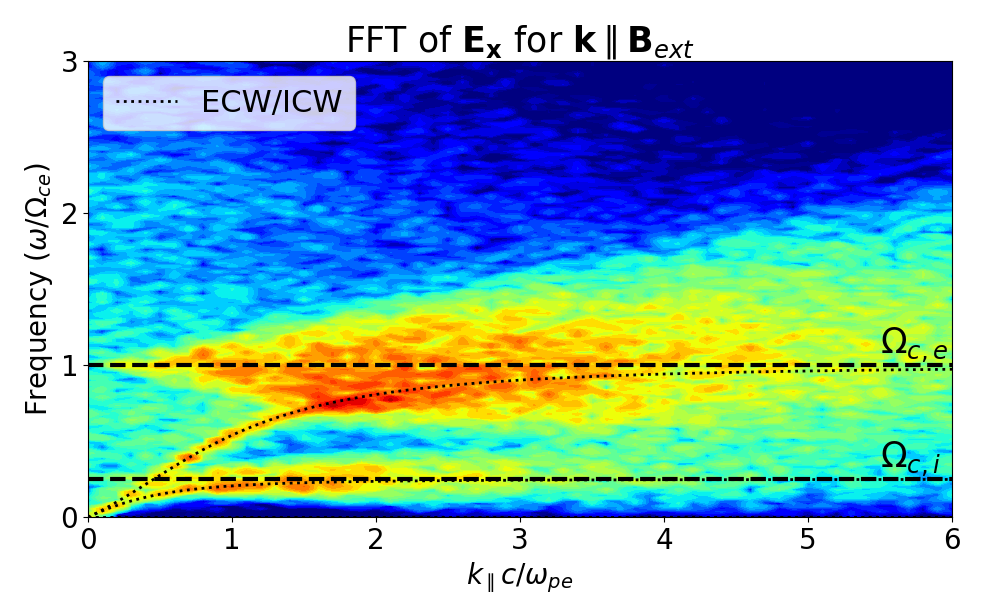}
        \caption{}
        \label{fig:kpar_double_Eperp_Ey}
    \end{subfigure}
    \hfill
    % Second row
    \begin{subfigure}[b]{0.67\textwidth}
        \centering
        \includegraphics[width=\linewidth, trim=15pt 0pt 20pt 0pt, clip]{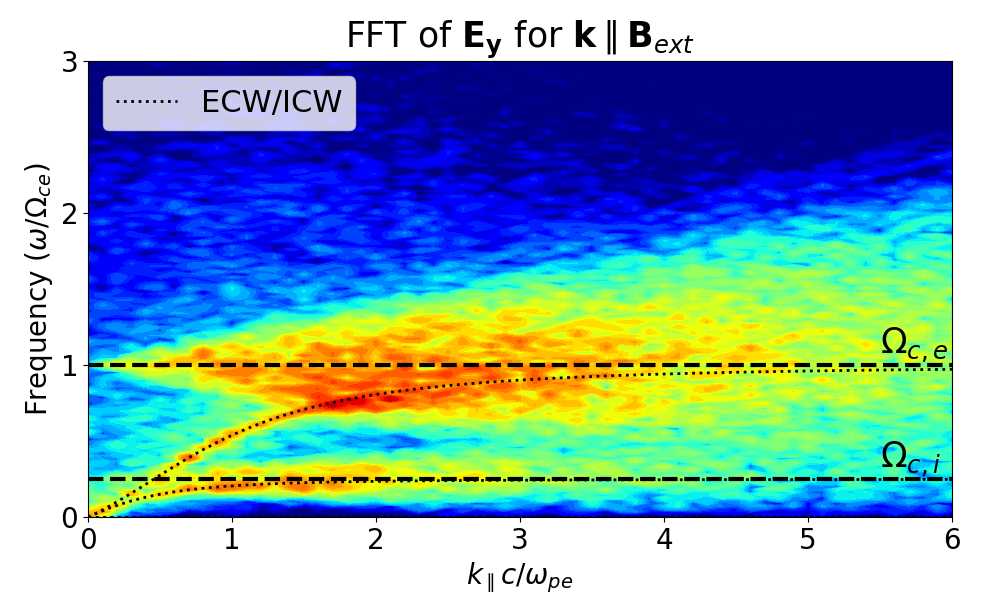}
        \caption{}
        \label{fig:kpar_double_Eperp_Ez}
    \end{subfigure}
    \hfill
    \begin{subfigure}[b]{0.67\textwidth}
        \centering
        \includegraphics[width=\linewidth, trim=15pt 0pt 20pt 0pt, clip]{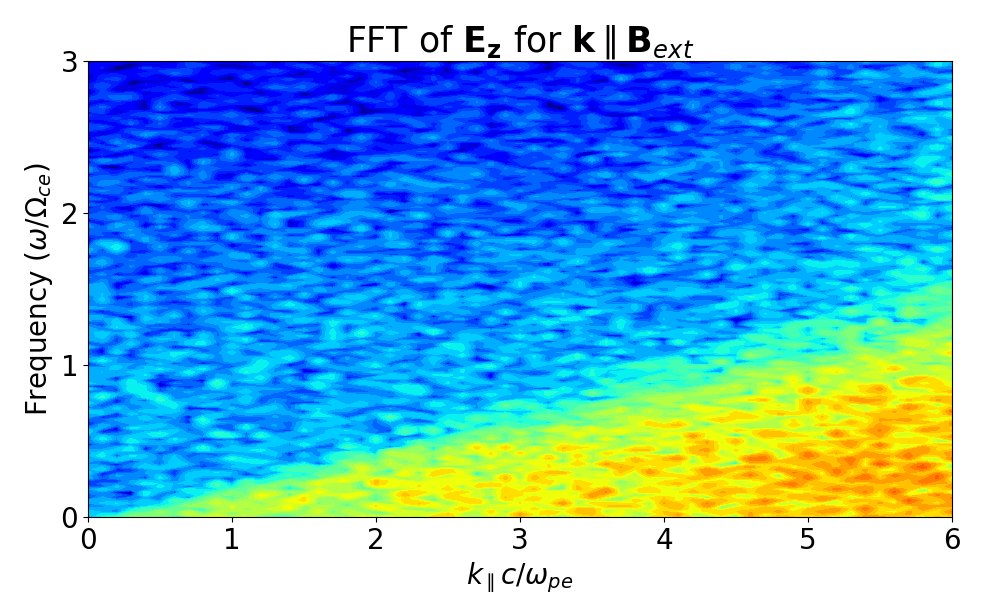}
        \caption{}
        \label{fig:kpar_double_Epar}
    \end{subfigure}
    \caption{Wave spectra of (a) $E_x$, (b) $E_y$ and (c) $E_z(E_\parallel)$ for a double-species electron-ion plasma with $v_{th,e} = 0.05c$ and $v_{th,i} = 0.025c$, for $\bk \parallel \bB_{ext}$. $E_x$ and $E_y$ waves are transverse and $E_z(E_\parallel)$ waves are longitudinal. In (a) and (b), the upper and lower dotted lines show the electron (ECW) and ion cyclotron waves (ICW) respectively, while the thick dashed lines show the respective cyclotron frequencies $\Omega_{c,e}$ and $\Omega_{c,i}$.}
    \label{fig:kpar_double_Eperp}
\end{figure}

A double-species simulation with kinetic electrons and ions both, is considered next. The number of particles per cell is 500 for each species. A reduced mass ratio of $m_i/m_e = 4$ is used. Hence, for an electron thermal velocity of $v_{th,e} = 0.05c$, the ion thermal velocity is given by $v_{th,i} = 0.025c$ to satisfy the requirement of thermal equilibrium. Using $B_{ext} = 1$, the $E_x$ and $E_y$ wave spectra obtained are shown in Figures \ref{fig:kpar_double_Eperp_Ey} and \ref{fig:kpar_double_Eperp_Ez}, respectively, and the $E_z(E_\parallel)$ spectrum is shown in Figure \ref{fig:kpar_double_Epar}. As in the electron-only simulation, the electron cyclotron wave (ECW) can be clearly seen. Additionally, we also see the ion cyclotron wave (ICW) now, with its analytical relation from \eqref{eq:coldplasma_par_L} and the $\Omega_{c,i}$ line superimposed on it. For this mode, the frequency asymptotes to the ion cyclotron frequency $\Omega_{c,i}$ as $k_\parallel \rightarrow \infty$. For the perpendicular field components, we observe the conical structure of the corresponding damped higher-order Alfv\'en-cyclotron modes, for both cyclotron frequencies. As expected from quasineutrality, the Langmuir wave and the electromagnetic L-mode and R-mode are missing in the wave spectra.

\subsection{Waves propagating $\perp$ to $\bB_{ext}$}\label{sec:sim_perpendicularwaves}

\subsubsection{Single-species plasma}\label{sec:sim_perpendicularwaves_single}
\begin{figure}
    \centering
    % First row
    \begin{subfigure}[b]{0.49\textwidth}
        \centering
        \includegraphics[width=\linewidth, trim=15pt 0pt 20pt 0pt, clip]{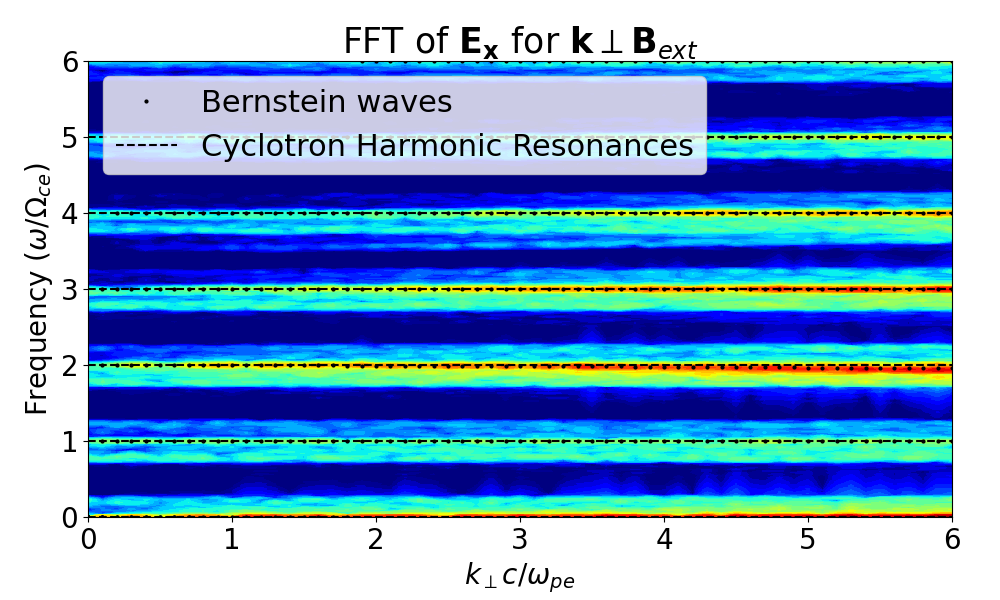}
        \caption{}
        \label{fig:kperp_single_Eperp_Ex}
    \end{subfigure}
    \hfill
    \begin{subfigure}[b]{0.49\textwidth}
        \centering
        \includegraphics[width=\linewidth, trim=15pt 0pt 20pt 0pt, clip]{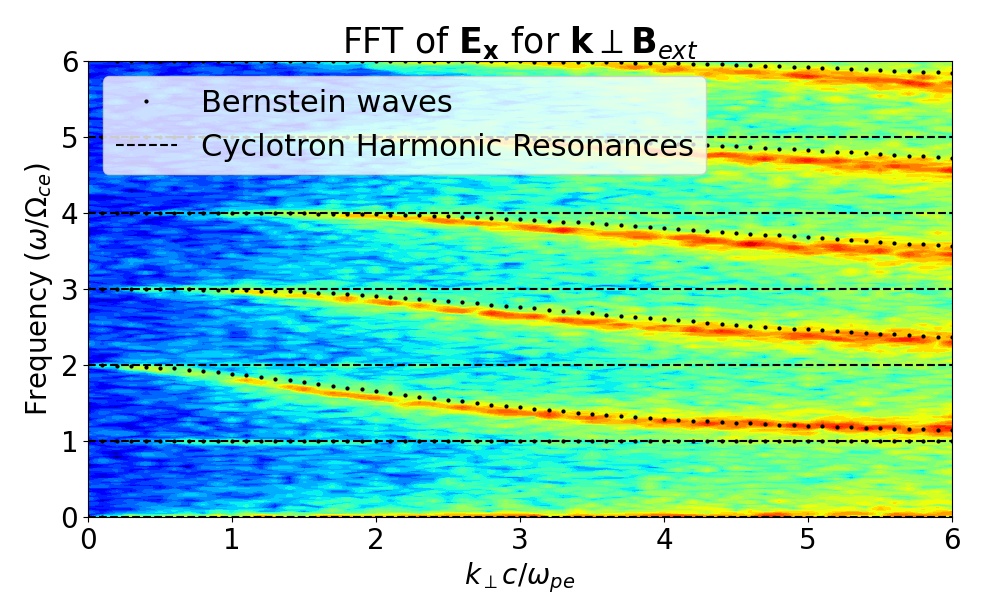}
        \caption{}
        \label{fig:kperp_single_Eperp_hot_Ex}
    \end{subfigure}
    \hfill
    \begin{subfigure}[b]{0.49\textwidth}
        \centering
        \includegraphics[width=\linewidth, trim=15pt 0pt 20pt 0pt, clip]{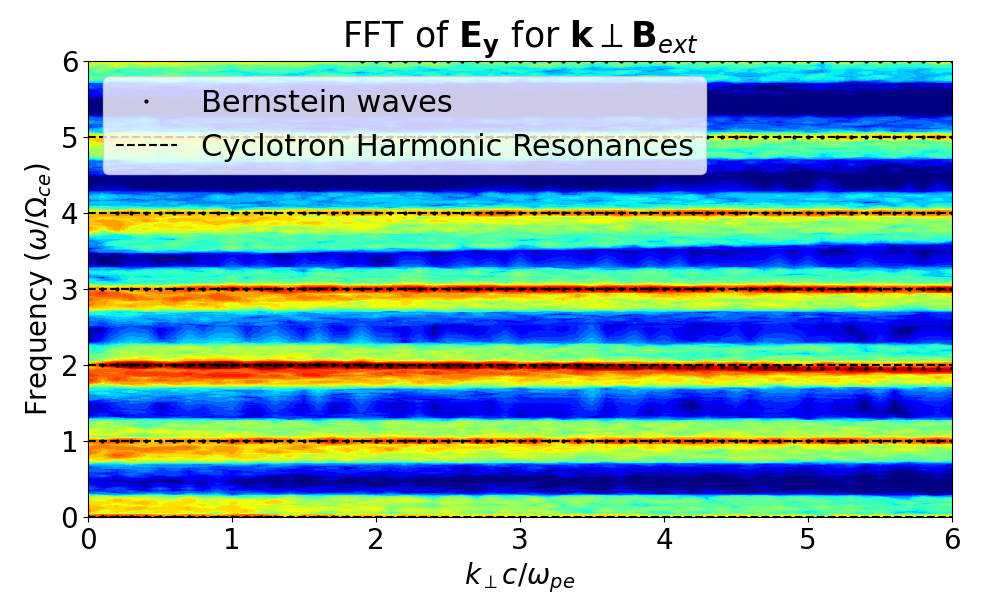}
        \caption{}
        \label{fig:kperp_single_Eperp_Ey}
    \end{subfigure}
    \hfill
    \begin{subfigure}[b]{0.49\textwidth}
        \centering
        \includegraphics[width=\linewidth, trim=15pt 0pt 20pt 0pt, clip]{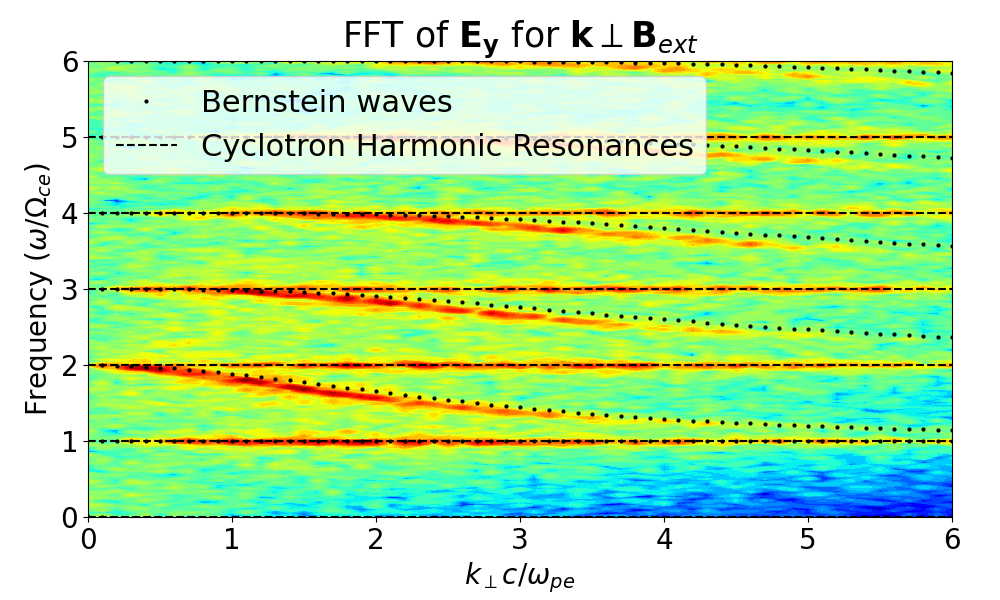}
        \caption{}
        \label{fig:kperp_single_Eperp_hot_Ey}
    \end{subfigure}
    \hfill
    \begin{subfigure}[b]{0.49\textwidth}
        \centering
        \includegraphics[width=\linewidth, trim=15pt 0pt 20pt 0pt, clip]{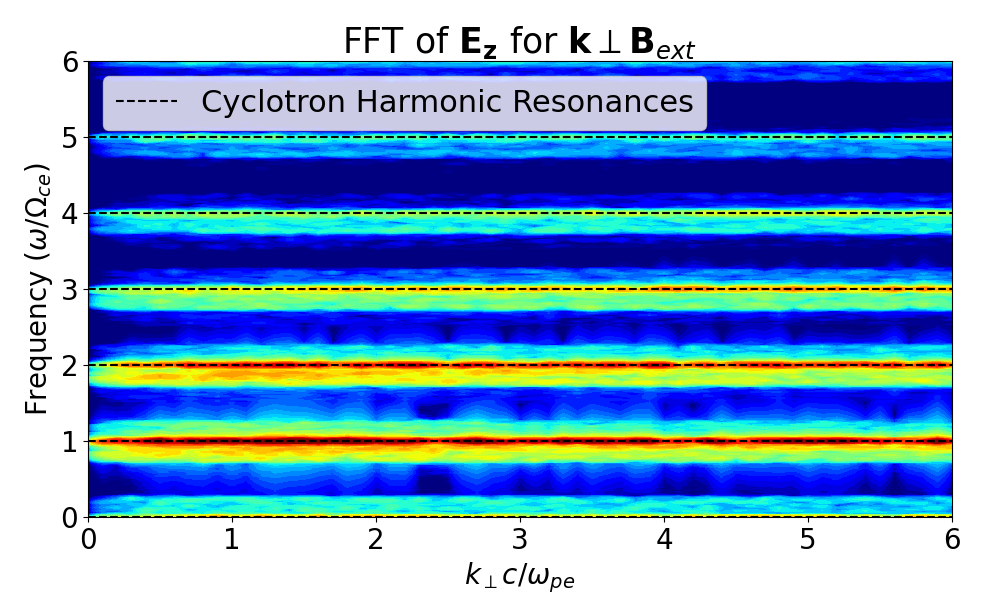}
        \caption{}
        \label{fig:kperp_single_Epar}
    \end{subfigure}
    \hfill
    \begin{subfigure}[b]{0.49\textwidth}
        \centering
        \includegraphics[width=\linewidth, trim=15pt 0pt 20pt 0pt, clip]{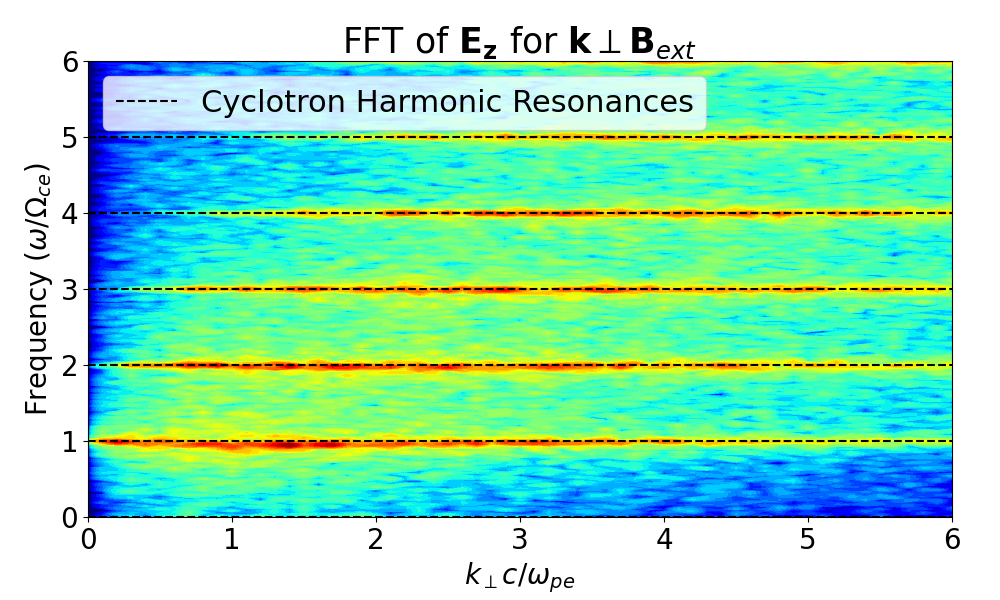}
        \caption{}
        \label{fig:kperp_single_Epar_hot}
    \end{subfigure}
    \caption{Wave spectra of (a) $E_x$, (c) $E_y$ and (e) $E_z(E_\parallel)$ for $v_{th,e} = 0.05c$, and of (b) $E_x$, (d) $E_y$ and (f) $E_z(E_\parallel)$ for $v_{th,e} = 0.5c$, for a single-species electron-only plasma, for $\bk \perp \bB_{ext}$. $E_x$ waves are longitudinal and $E_y$ and $E_z(E_\parallel)$ waves are transverse. The dotted lines show the Bernstein waves while the dashed lines show the cyclotron harmonic resonances. These are almost overlapping in (a) and (c) and are distinct in (b) and (d).}
    \label{fig:kperp_single}
\end{figure}

We now turn our attention to waves propagating in the direction perpendicular to that of the background magnetic field. We first simulate an electron-only plasma, with stationary ions in the background. A total of 1000 particles per cell are used for the electron species. The thermal velocity of the electrons is $v_{th,e} = 0.05c$. The 256-cell resolution is now along a direction perpendicular to the background magnetic field, namely the $x$-direction, as this is the direction of propagation of the waves studied. A normalized background magnetic field of magnitude $B_{ext} = 1$ is used. The wave spectrum for the $E_z(E_\parallel)$ field is shown in Figure \ref{fig:kperp_single_Epar}. We see here the resonant modes concentrated at the harmonics of the electron cyclotron frequency, as described above in Section \ref{sec:perpendicularwaves}. The wave spectra for the $E_x$ and $E_y$ fields are shown in Figures \ref{fig:kperp_single_Eperp_Ex} and \ref{fig:kperp_single_Eperp_Ey}, respectively. The waves contained in these spectra are the Bernstein waves described above in Section \ref{sec:perpendicularwaves}, as well as the cyclotron harmonic resonances. However, due to the Bernstein waves also being concentrated around the harmonics, it is impossible to distinguish between the two. The theoretical locations of these harmonic resonances are marked with dashed lines. Due to quasineutrality, the ordinary or O-mode is missing in $E_z(E_\parallel)$ spectrum in Figure \ref{fig:kperp_single_Epar}, while the X-mode is missing in the $E_x$ and $E_y$ spectra in Figures \ref{fig:kperp_single_Eperp_Ex} and \ref{fig:kperp_single_Eperp_Ey}. These can be seen in the work by \citet{meng2025}.

\begin{figure}
    \centering
    % First row
    \begin{subfigure}[b]{0.49\textwidth}
        \centering
        \includegraphics[width=\linewidth, trim=15pt 0pt 20pt 0pt, clip]{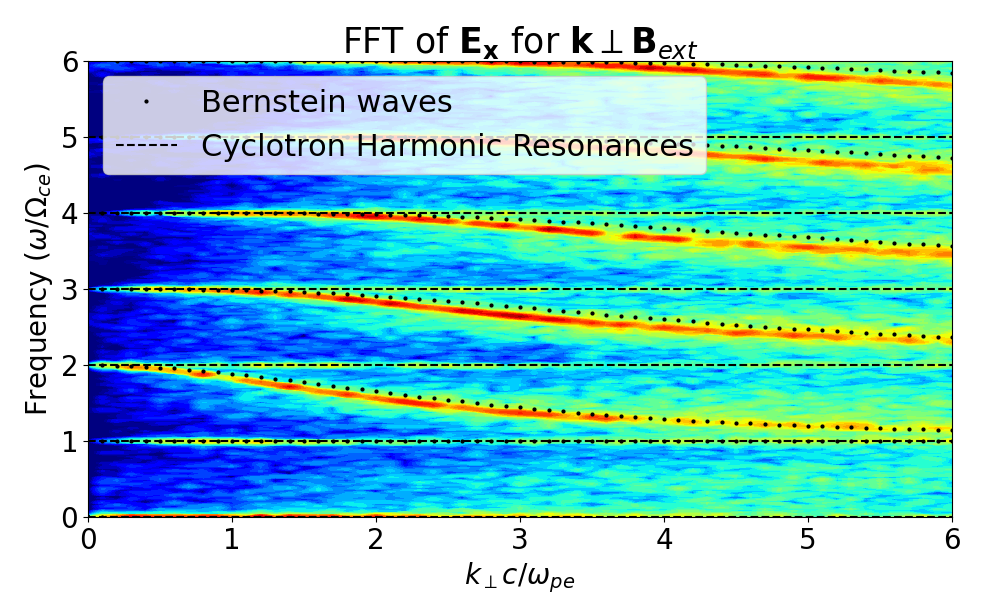}
        \caption{}
        \label{fig:kperp_single_Eperp_hot_Ex_field}
    \end{subfigure} \\
    \begin{subfigure}[b]{0.49\textwidth}
        \centering
        \includegraphics[width=\linewidth, trim=15pt 0pt 20pt 0pt, clip]{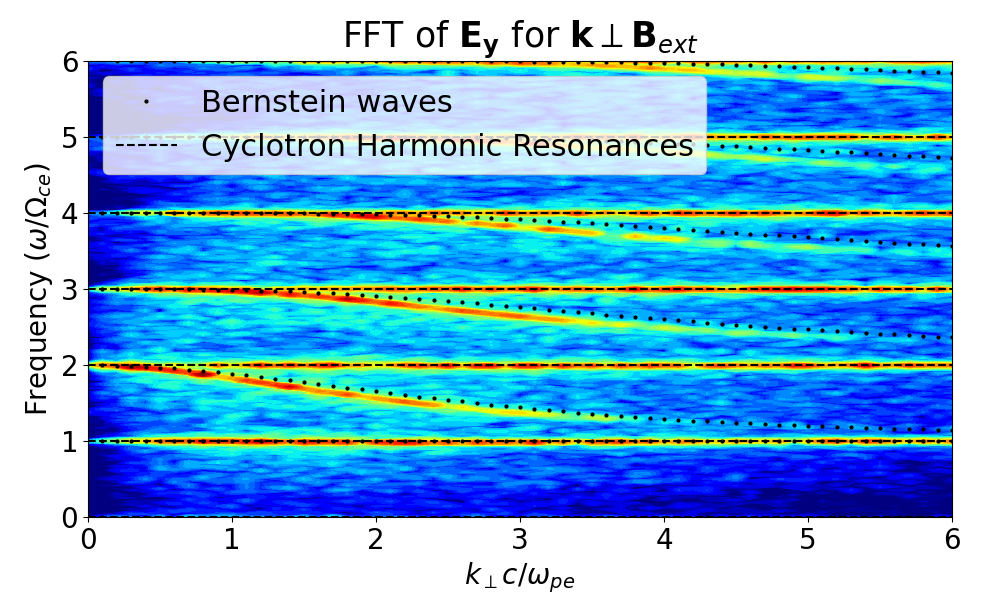}
        \caption{}
        \label{fig:kperp_single_Eperp_hot_Ey_field}
    \end{subfigure} \\
    \begin{subfigure}[b]{0.49\textwidth}
        \centering
        \includegraphics[width=\linewidth, trim=15pt 0pt 20pt 0pt, clip]{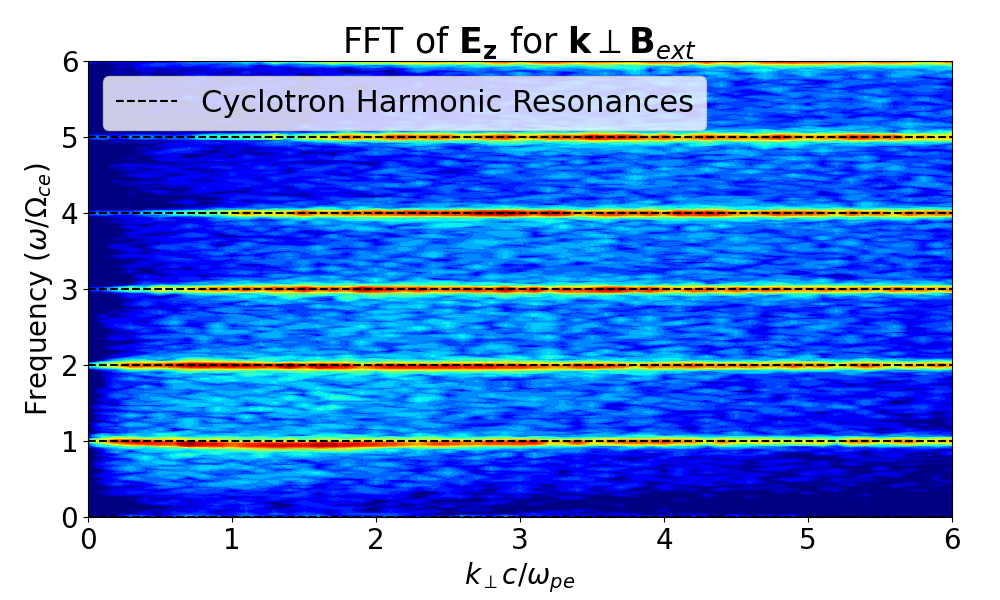}
        \caption{}
        \label{fig:kperp_single_Epar_hot_field}
    \end{subfigure}
    \caption{Wave spectra of (a) $E_x$, (b) $E_y$ and (v) $E_z(E_\parallel)$ for $v_{th,e} = 0.5c$, for a single-species electron-only plasma, for $\bk \perp \bB_{ext}$, using the computationally cheaper approach wherein equation \eqref{eq:efieldeqn_disc_rhofieldcorr} is used to obtain the electric field. $E_x$ waves are longitudinal and $E_y$ and $E_z(E_\parallel)$ waves are transverse. The dotted lines show the Bernstein waves while the dashed lines show the cyclotron harmonic resonances.}
    \label{fig:kperp_single_fieldrho}
\end{figure}

This case was also simulated for an electron thermal velocity of $v_{th,e} = 0.5c$, to observe the changes in the behavior of Bernstein waves in hotter plasmas. The corresponding wave spectra for the $E_x$, $E_y$ and $E_z(E_\parallel)$ fields for this simulation are shown in Figures \ref{fig:kperp_single_Eperp_hot_Ex}, \ref{fig:kperp_single_Eperp_hot_Ey} and \ref{fig:kperp_single_Epar_hot}, respectively. We observe clear shifts in the behavior of the Bernstein waves, wherein the curves starting at $k_{\perp}=0$ approach a lower harmonic at high values of $k_\perp$. The Bernstein waves can now be seen distinctly along with the harmonic resonances. Approximate analytical solutions of the Bernstein waves are also shown in these plots as dotted lines. The resonances are observed to be much weaker in the $E_x$ spectrum, with only the $n=1$ harmonic clearly visible and the $n=2$ harmonic barely visible after a closer look. The $n > 2$ harmonics are too weak to be detected. This case with the hotter plasma was also simulated using the alternative approach wherein the electric field is obtained using equation \eqref{eq:efieldeqn_disc_rhofieldcorr}. The spectra obtained using this approach are shown in Figure \ref{fig:kperp_single_fieldrho}. The results show a very strong match with those in Figures \ref{fig:kperp_single_Eperp_hot_Ex}, \ref{fig:kperp_single_Eperp_hot_Ey} and \ref{fig:kperp_single_Epar_hot}, although with noticeably lower noise levels. In terms of savings in the computational cost, the time required for this simulation was $\sim25\%$ lower, as compared to the approach that used equation \eqref{eq:efieldeqn_disc} to obtain the electric field.

\subsubsection{Double-species plasma}\label{sec:sim_perpendicularwaves_double}
\begin{figure}
    \centering
    % First row
    \begin{subfigure}[b]{0.49\textwidth}
        \centering
        \includegraphics[width=\linewidth, trim=15pt 0pt 20pt 0pt, clip]{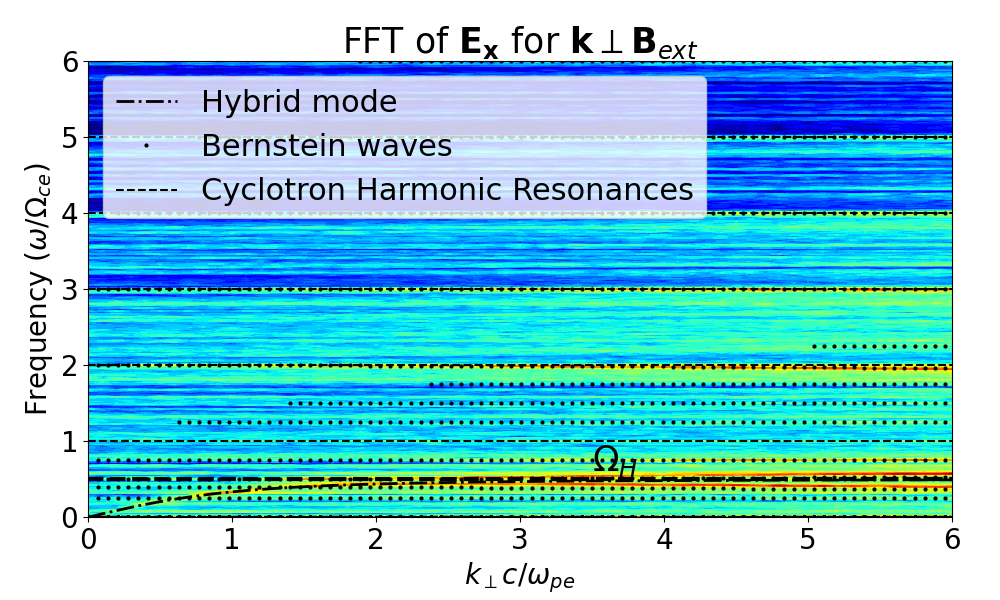}
        \caption{}
        \label{fig:kperp_double_Eperp_Ex}
    \end{subfigure}
    \hfill
    \begin{subfigure}[b]{0.49\textwidth}
        \centering
        \includegraphics[width=\linewidth, trim=15pt 0pt 20pt 0pt, clip]{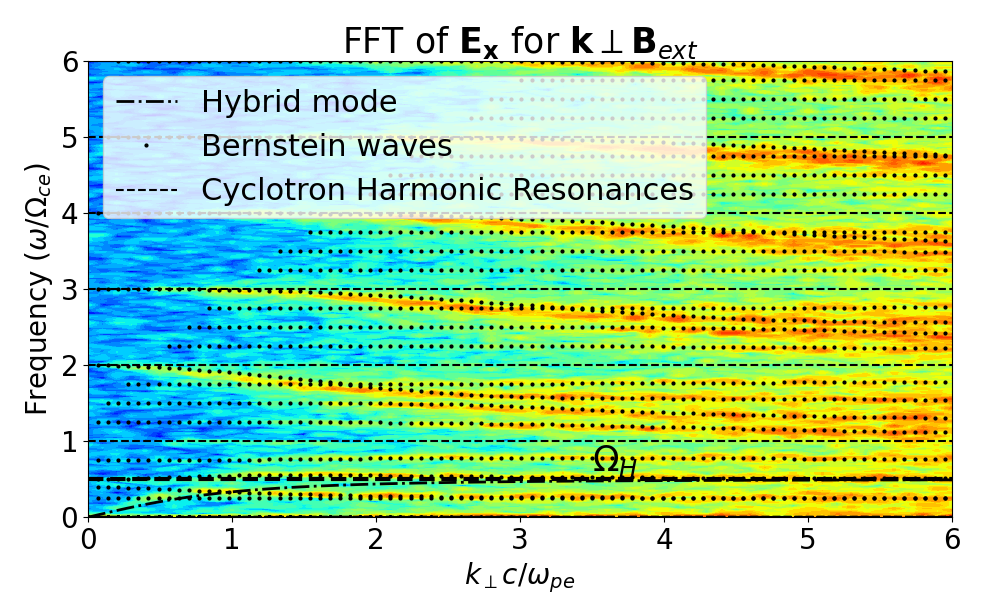}
        \caption{}
        \label{fig:kperp_double_Eperp_hot_Ex}
    \end{subfigure}
    \hfill
    \begin{subfigure}[b]{0.49\textwidth}
        \centering
        \includegraphics[width=\linewidth, trim=15pt 0pt 20pt 0pt, clip]{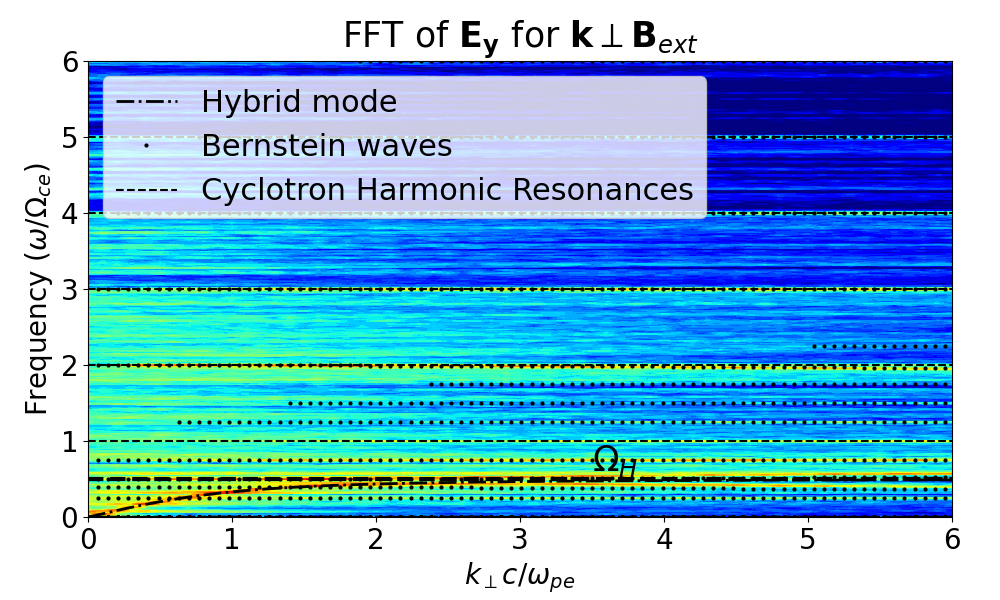}
        \caption{}
        \label{fig:kperp_double_Eperp_Ey}
    \end{subfigure}
    \hfill
    \begin{subfigure}[b]{0.49\textwidth}
        \centering
        \includegraphics[width=\linewidth, trim=15pt 0pt 20pt 0pt, clip]{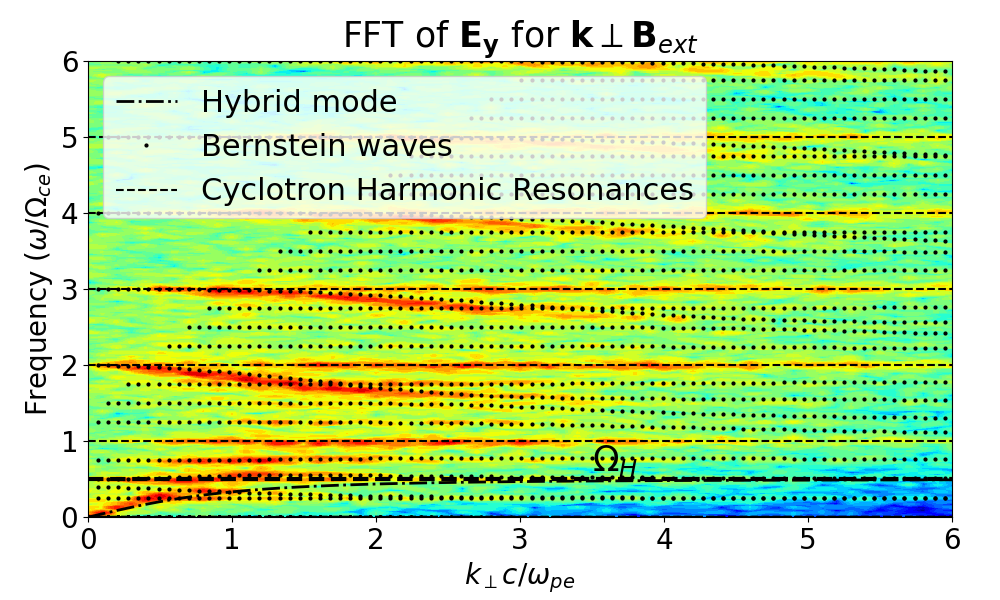}
        \caption{}
        \label{fig:kperp_double_Eperp_hot_Ey}
    \end{subfigure}
    \hfill
    \begin{subfigure}[b]{0.49\textwidth}
        \centering
        \includegraphics[width=\linewidth, trim=15pt 0pt 20pt 0pt, clip]{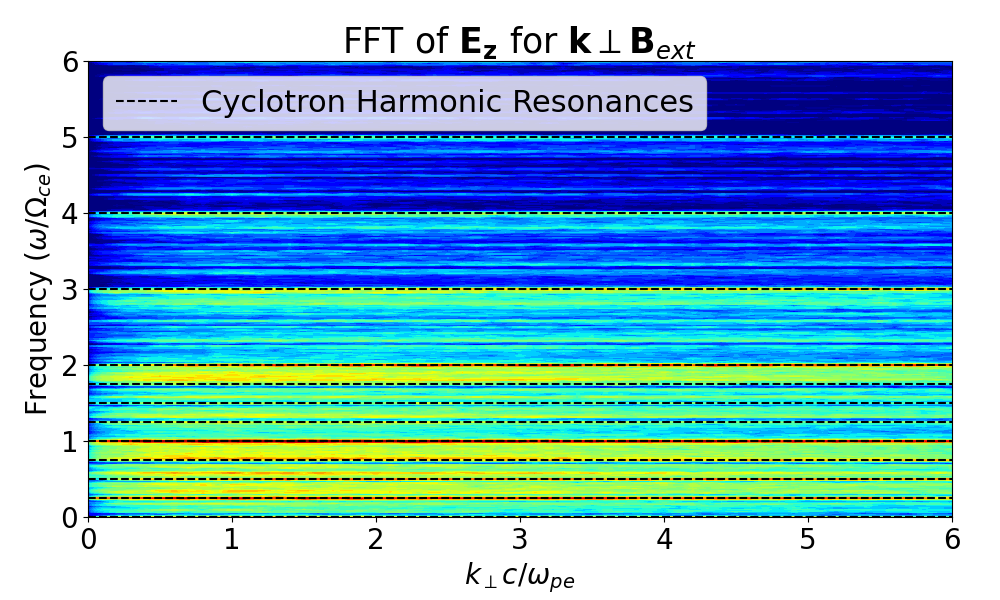}
        \caption{}
        \label{fig:kperp_double_Epar}
    \end{subfigure}
    \hfill
    \begin{subfigure}[b]{0.49\textwidth}
        \centering
        \includegraphics[width=\linewidth, trim=15pt 0pt 20pt 0pt, clip]{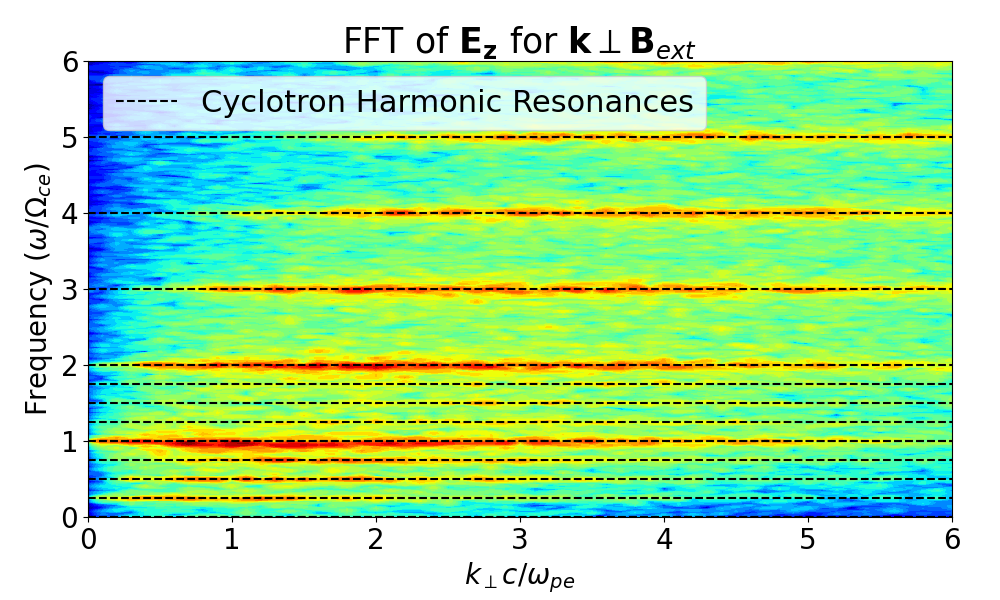}
        \caption{}
        \label{fig:kperp_double_Epar_hot}
    \end{subfigure}
    \caption{Wave spectra of (a) $E_x$, (c) $E_y$ and (e) $E_z(E_\parallel)$ for $v_{th,e} = 0.05c$ and $v_{th,i} = 0.025c$, and of (b) $E_x$, (d) $E_y$ and (f) $E_z(E_\parallel)$ for $v_{th,e} = 0.5c$ and $v_{th,i} = 0.25c$, for a double-species electron-ion plasma, for $\bk \perp \bB_{ext}$. $E_x$ waves are longitudinal and $E_y$ and $E_z(E_\parallel)$ waves are transverse. The horizontal dashed lines show both the electron and ion cyclotron harmonic resonances. Only the first eight modes of ion cyclotron harmonics are shown for clarity. In (a), (b), (c) and (d), the dotted lines represent the electron and ion Bernstein waves, while the dash-dot line shows the hybrid wave. The thicker dashed line represents the hybrid frequency.}
    \label{fig:kperp_double}
\end{figure}

\begin{figure}
    \centering
    % First row
    \begin{subfigure}[b]{0.49\textwidth}
        \centering
        \includegraphics[width=\linewidth, trim=15pt 0pt 20pt 0pt, clip]{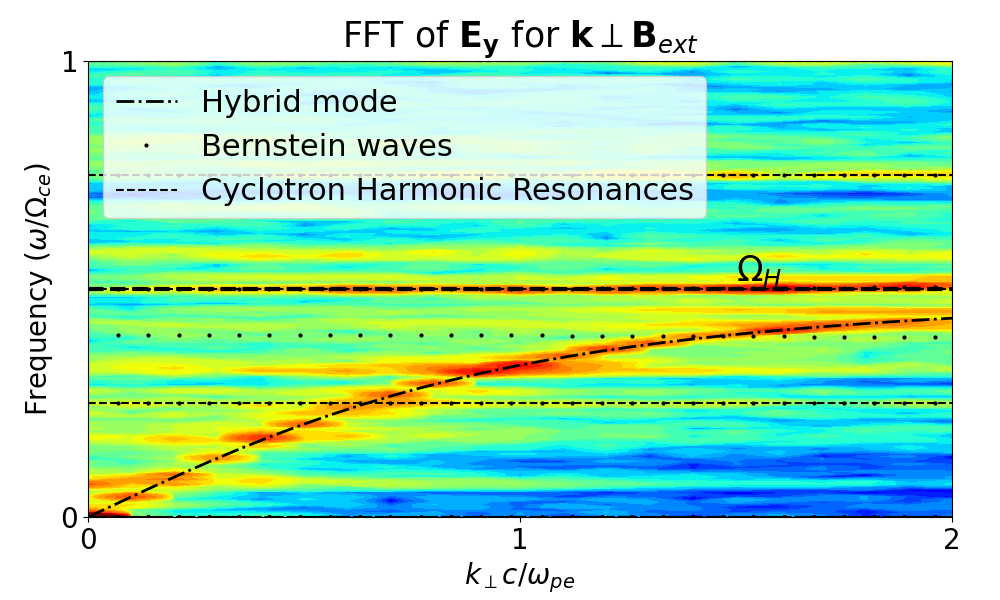}
        \caption{}
        \label{fig:kperp_double_Eperp_Ey_mag}
    \end{subfigure}
    \hfill
    \begin{subfigure}[b]{0.49\textwidth}
        \centering
        \includegraphics[width=\linewidth, trim=15pt 0pt 20pt 0pt, clip]{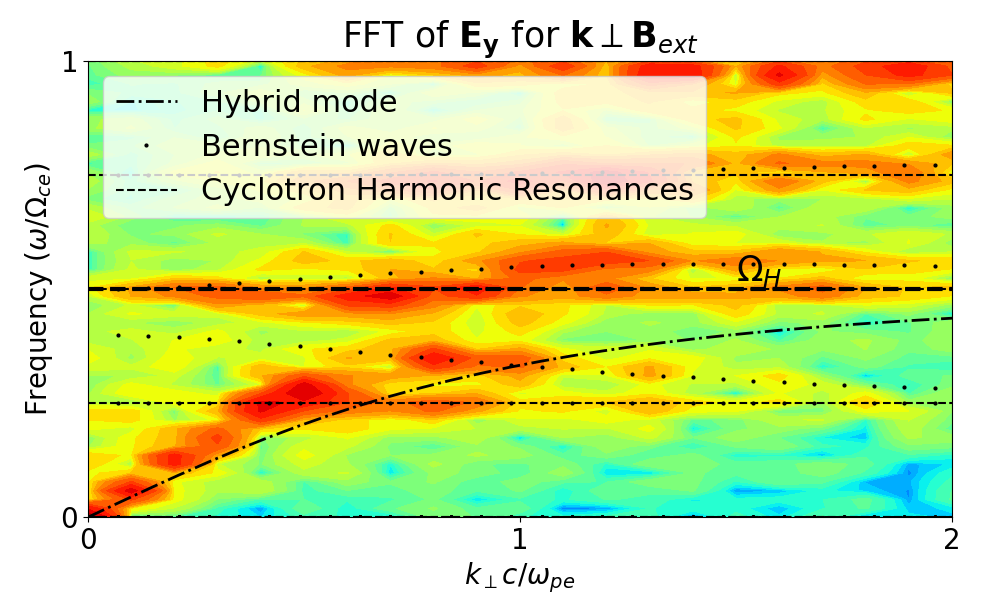}
        \caption{}
        \label{fig:kperp_double_Eperp_hot_Ey_mag}
    \end{subfigure}
\caption{Magnified view of the hybrid wave observed in Figures \ref{fig:kperp_double_Eperp_Ey} (a) and \ref{fig:kperp_double_Eperp_hot_Ey} (b). The dash-dot line shows the hybrid wave and the thick dashed line represents the hybrid frequency.}
\label{fig:kperp_double_Eperp_mag}
\end{figure}

As a final case for validating wave spectra, a double-species electron-ion plasma is considered. For each species, we use 500 particles per cell. The reduced mass ratio and the electron and ion thermal velocities are given by $m_i/m_e = 4$, $v_{th,e} = 0.05c$, and $v_{th,i} = 0.025c$. The normalized magnetic field is again $B_{ext} = 1$. The wave spectra obtained for the $E_x$, $E_y$ and $E_z(E_\parallel)$ fields for this simulation are shown in Figures \ref{fig:kperp_double_Eperp_Ex}, \ref{fig:kperp_double_Eperp_Ey} and \ref{fig:kperp_double_Epar}, respectively. The corresponding spectra obtained for a hotter plasma with thermal velocities $v_{th,e} = 0.5c$, and $v_{th,i} = 0.25c$ are shown in Figures \ref{fig:kperp_double_Eperp_hot_Ex}, \ref{fig:kperp_double_Eperp_hot_Ey} and \ref{fig:kperp_double_Epar_hot}, respectively. Besides the waves obtained in the electron-only simulation, we now also see other waves that can be attributed to the presence of the second species i.e. ions. In Figures \ref{fig:kperp_double_Epar} and \ref{fig:kperp_double_Epar_hot}, we notice the resonant modes concentrated at harmonics of the ion cyclotron frequency. This frequency is 1/4 times the electron cyclotron frequency for the reduced mass ratio of $m_i/m_e = 4$ used here. These modes are therefore concentrated at relatively lower frequencies on the $k_\perp$-$\omega$ plot and much weaker at higher frequencies as can be seen in Figure \ref{fig:kperp_double_Epar}. They are almost indiscernible in Figure \ref{fig:kperp_double_Epar_hot} for higher frequencies, for the hotter plasma. For the sake of clarity of the figures, only the first few harmonics of the ion cyclotron frequency are marked with dashed lines, along with those of the electron cyclotron frequency. These ion cyclotron harmonic dashed lines are not shown in other figures to avoid clutter. In Figures \ref{fig:kperp_double_Eperp_Ex} and \ref{fig:kperp_double_Eperp_Ey}, we now also see concentrations of lines at harmonics of the ion cyclotron frequency. These are ion cyclotron resonances and ion Bernstein waves overlapping with each other. Again, due to quasineutrality, the O-mode and X-modes are missing from the wave spectra. The full Vlasov-Maxwell system would also show a second hybrid wave in this case, on account of the second specied being present. This wave can be observed in the work by \citet{meng2025}. In \ref{fig:kperp_double_Eperp_hot_Ex} and \ref{fig:kperp_double_Eperp_hot_Ey}, for the hotter plasma, we distinctly observe ion Bernstein waves as their curves deviate from the harmonics. Approximate analytical solutions for ion Bernstein waves are also superimposed on top, with some of them intersecting those of the electron Bernstein waves. Additionally, the hybrid wave arising from the presence of a second species is also seen in Figures \ref{fig:kperp_double_Eperp_hot_Ex} and \ref{fig:kperp_double_Eperp_hot_Ey}, with the hybrid wave analytical solution and $\Omega_H$ line superimposed on in. Magnified views of the hybrid waves in Figures \ref{fig:kperp_double_Eperp_Ey} and \ref{fig:kperp_double_Eperp_hot_Ey} are shown in Figure \ref{fig:kperp_double_Eperp_mag}. While the analytical solution of the hybrid wave shows an excellent match with the numerical result in Figure \ref{fig:kperp_double_Eperp_Ey_mag}, a significant difference between the two is seen Figure \ref{fig:kperp_double_Eperp_hot_Ey_mag}. This is because the analytical result of equation \eqref{eq:cold_plasma_perp} is valid only in the cold plasma limit, and does not account for any deviations arising from high-temperature effects. These effects are much more pronounced for the hotter plasma with $v_{th,e} = 0.5c$. %\nmn{This case was also run for various values of particle count per cell, to observe the effect of particle count on density error. Particle per cell counts between 250 and 1250 are considered, with an increment of 250 between runs. The percentage error $|\rho/\rho_i| \times 100$ where $\rho_i$ is the ion charge density while $\rho = \rho_e + \rho_i$ is the total charge density which should ideally be zero according to the quasineutrality condition. This error is plotted against the particle count per cell in Figure \ref{fig:Rhoerror}. Its slope shows a good match with the analytical slope given by \citet{birdsall2018plasma}.}

%\begin{figure}
%\centering
%    \begin{subfigure}[b]{\textwidth}
%        \centering
%        \includegraphics[width=0.99\linewidth, trim=0pt 0pt 0pt 0pt, clip]{./Figs/Rho_percentage_error.png}
%        \caption{}
%        \label{fig:Damping_Ey}
%    \end{subfigure}
%\caption{\nmn{Density percentage error i.e. $|\rho/\rho_i|\times 100$ for the double-species electron-ion plasma simulation with $v_{th,e} = 0.5c$ and $v_{th,i} = 0.25c$, for $\bk \perp \bB_{ext}$. The particle count per cell for each species considered is 250, 500, 750, 1000 and 1250. The blue graph shows the result obtained from the numerical simulation, while the dashed red graph shows the expected analytical result predicted by %\citet{birdsall2018plasma}.}}
%\label{fig:Rhoerror}
%\end{figure}

\subsection{Damping of waves}\label{sec:sim_damping}

\begin{figure}
\centering
    % First row
    \begin{subfigure}[b]{\textwidth}
        \centering
        \includegraphics[width=0.99\linewidth, trim=0pt 0pt 0pt 0pt, clip]{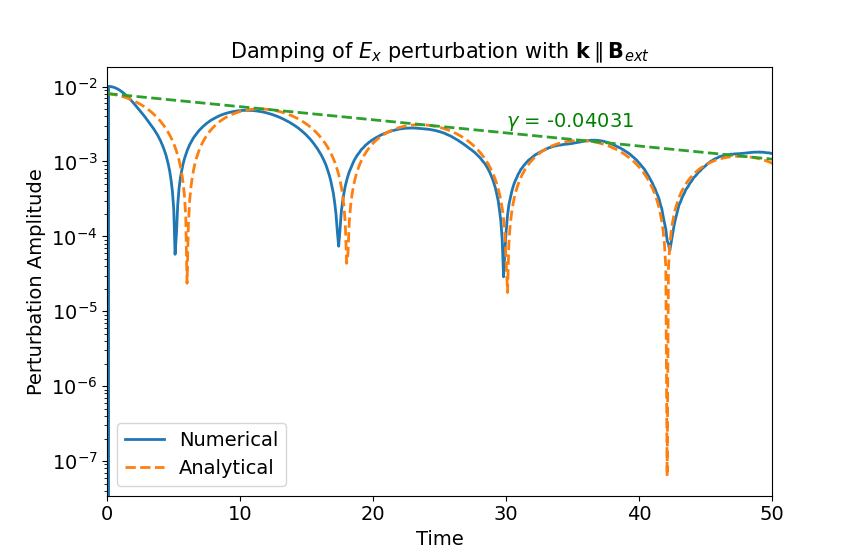}
        \caption{}
        \label{fig:Damping_Ex}
    \end{subfigure}
    \hfill
    \begin{subfigure}[b]{\textwidth}
        \centering
        \includegraphics[width=0.99\linewidth, trim=0pt 0pt 0pt 0pt, clip]{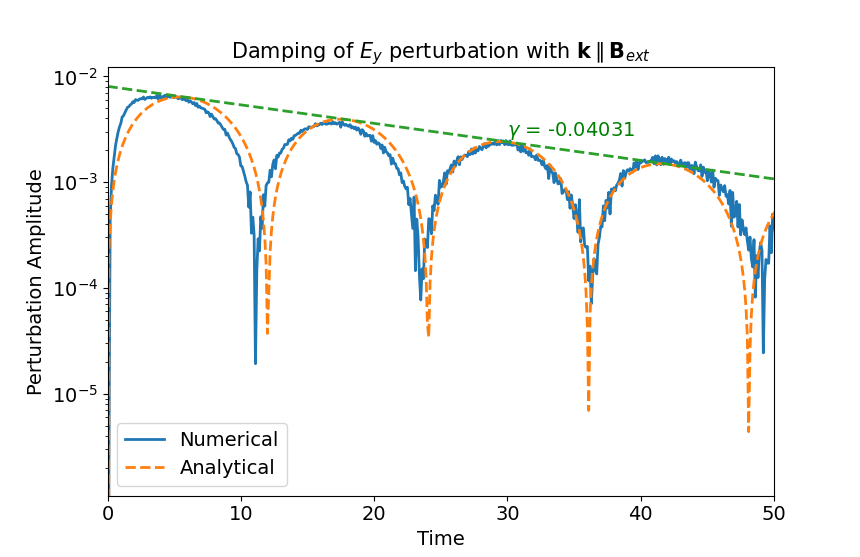}
        \caption{}
        \label{fig:Damping_Ey}
    \end{subfigure}
\caption{Damping of $E_x$ (a) and $E_y$ (b) perturbations for $k_{\parallel} = 0.7$ in electron-only plasma with $v_{th,e} = 0.5c$. The blue graph shows the result obtained from the numerical simulation, while the dashed orange graph shows the expected analytical result. The dashed green line shows the analytical damping rate.}
\label{fig:Damping}
\end{figure}

The final test considered is a cyclotron wave-damping simulation wherein the numerical damping rate of a single sinusoidal perturbation of a particular wavenumber is compared with its analytical value obtained from the dispersion relation. We consider a single-species electron-only plasma with stationary ions in a normalized background magnetic field of $B_{ext} = 1$. Without loss of generality, we assume the $\bB_{ext}$ direction to be the $z$-direction. A wave propagating in the direction parallel to the background magnetic field is excited. For this purpose, a sinusoidal velocity perturbation given by
\begin{equation}\label{eq:vel_perturbation}
  v_y = 0.03 \sin (k_{\parallel} z),
\end{equation}
is used as an initial condition for the average bulk velocity of the plasma. Here, a parallel wavenumber of $k_{\parallel} = 0.7$ is used. Significant damping rates require higher plasma temperatures. The electron thermal velocity is therefore taken to be $v_{th,e} = 0.5c$. A computational grid of $128\times8\times8$ cells is used. As the excited wave is propagating in the direction parallel to $\bB_{ext}$, the higher resolution of 128 cells is along this direction. Damping of waves is a higher-order process and relatively difficult to capture in PIC methods. To accurately capture this process, we use 5000 particles per cell. The perturbation given in equation \eqref{eq:vel_perturbation} creates sinusoidal perturbations in the electric fields in the perpendicular $x$- and $y$- directions. These can be calculated from equation \eqref{eq:qnEeqn}. The dispersion properties of these perturbations are obtained from solving the warm plasma dispersion relation $((\epsilon_{xx}-n^2)(\epsilon_{yy}-n^2)-\epsilon_{xy}\epsilon_{yx}) = 0$ for the waves propagating parallel to the background magnetic field, where $\epsilon_{xx}$, $\epsilon_{yy}$, $\epsilon_{xy}$ and $\epsilon_{yx}$ are given in equations \eqref{eq:epsilonxx_par} and \eqref{eq:epsilonxy_par}. For the wavenumber $k_{\parallel} = 0.7$, the corresponding $\omega$ obtained from solving this equation is $\omega(k_\parallel = 0.7) \approx 0.2612 - 0.04031j$. Here, the real part is the angular speed and the imaginary part is the damping rate. The analytical time-evolutions of the $E_x$ and $E_y$ perturbation amplitudes obtained from $\omega(k_\parallel = 0.7)$ are plotted alongside the results from the numerical simulation in Figures \ref{fig:Damping_Ex} and \ref{fig:Damping_Ey}, respectively.

%%%%%%%%%%%%%%%%%%%%%%%%%%%%%%%%%%%%%%%%%%%%%%%%%%%%%%%%%

\subsection{Parallelization strategy and scalability}\label{sec:scaling}

\nmn{The implementation used in this work is based on the \texttt{AMReX} software framework \citep{zhang2021amrex}, which provides support for block-structured parallel simulations using distributed-memory parallelism via MPI. The computational domain is decomposed into grid patches that are distributed across MPI ranks using the load-balancing mechanisms provided by \texttt{AMReX}. Particle management, inter-process communication, and particle redistribution are handled through the framework's built-in parallel infrastructure. Linear systems arising from the numerical model are solved using the \texttt{HYPRE} library through the interfaces provided by \texttt{AMReX}.}

As expected in most PIC codes, the dominant portion of the overall computational cost arises from particle loops. The key particle operations for this model include pushing particles using electromagnetic fields calculated at particle positions, and particle-to-mesh depositions to calculate particle-based quantities defined on the grid spaces. These particle-based quantities include charge density, current density, the Lorentz force and stress tensor terms on the right hand side of the curl-curl equation \eqref{eq:efieldeqn_disc}, the particle contributions to the charge matrix on the left hand side of the curl-curl equation \eqref{eq:efieldeqn_disc}, and the particle contributions to the `divergence-corrector' potential matrix on the left hand side of equation \eqref{eq:divJstar_psistar_disc}. For a double-species simulation using 500 particles per cell, the computational costs of the costliest operations as percentages of the overall cost in descending order are:
\begin{enumerate}
    \item Calculating particle contributions to the charge matrix on the left hand side of the curl-curl equation \eqref{eq:efieldeqn_disc} and building matrix $\approx 40 \%$.
    \item Calculating particle contributions to the `divergence-corrector' potential matrix on the left hand side of equation \eqref{eq:divJstar_psistar_disc} and building matrix $\approx 40 \%$.
    \item Pushing particles using electromagnetic fields $\approx 3 \%$.
    \item Solving linear systems in equations \eqref{eq:laplaceA_disc}, \eqref{eq:efieldeqn_disc} and \eqref{eq:divJstar_psistar_disc} $\approx 3 \%$.
    \item Calculating charge density, current density, the Lorentz force and stress tensor terms on the right hand side of the curl-curl equation \eqref{eq:efieldeqn_disc} $\approx 3 \%$.
\end{enumerate}
With increasing number of particles per cell, the overall percentage of the cost of solving the linear systems reduces even further. While these percentages vary slightly, depending on the individual problem parameters, they are a reliable representation of the relative computational costs for all the problems considered in this work. Thus, overall performance and scalability are primarily governed by the efficiency of particle operations.

\begin{figure}
\centering
    % First row
    \begin{subfigure}[b]{0.49\textwidth}
        \centering
        \includegraphics[width=\linewidth, trim=0pt 0pt 0pt 0pt, clip]{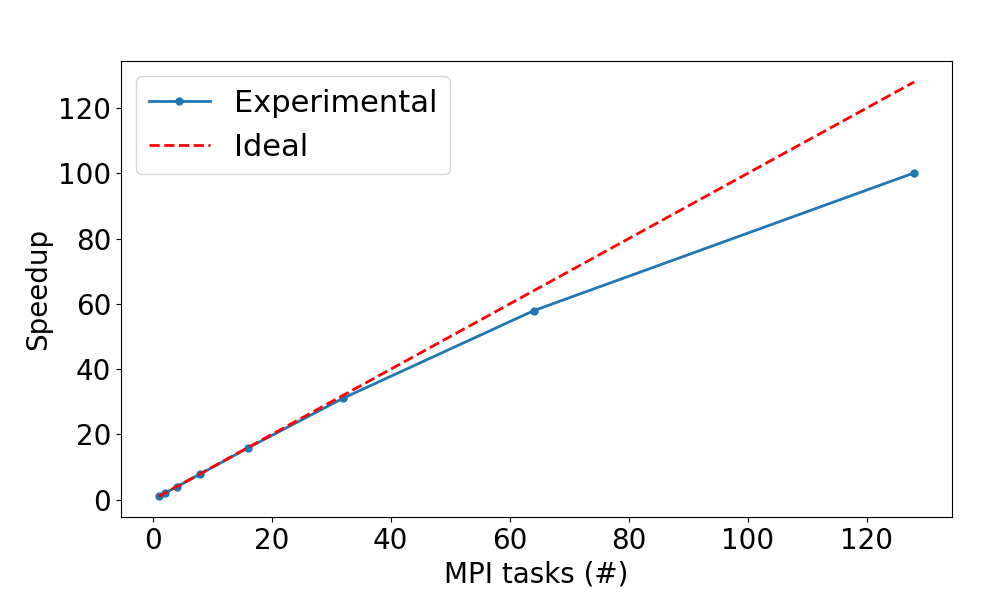}
        \caption{}
        \label{fig:StrongScaling}
    \end{subfigure}
    \hfill
    \begin{subfigure}[b]{0.49\textwidth}
        \centering
        \includegraphics[width=\linewidth, trim=0pt 0pt 0pt 0pt, clip]{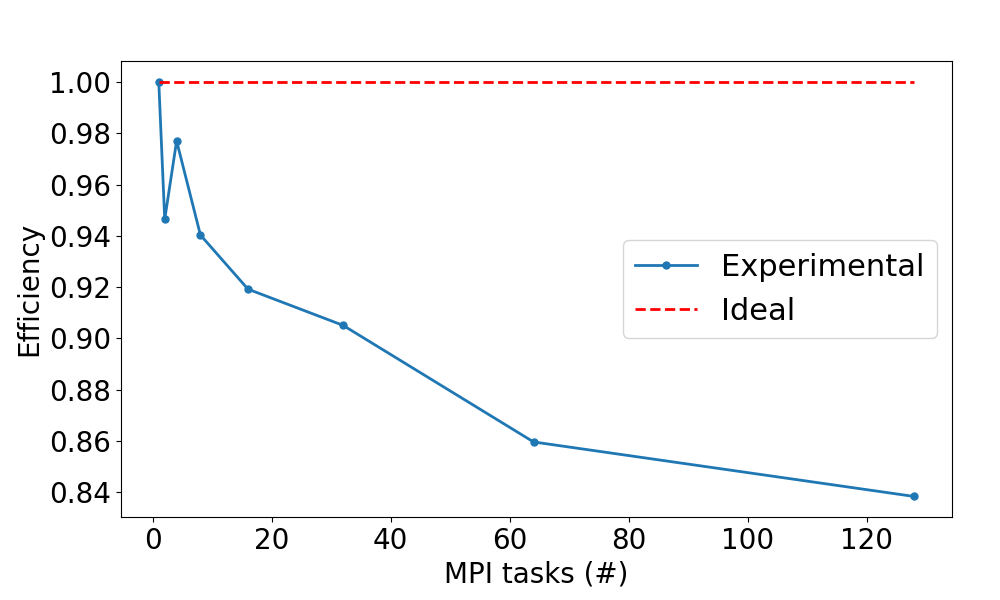}
        \caption{}
        \label{fig:WeakScaling}
    \end{subfigure}
\caption{Strong scaling performance showing speedup versus number of MPI tasks, keeping problem size constant (a), and weak scaling performance showing efficiency as number of MPI tasks increases in proportion to problem size (b).}
\label{fig:Damping}
\end{figure}

Strong and weak scaling studies were conducted to evaluate the parallel performance of the code. The case with the perpendicular wave propagation spectrum for a double-species plasma with $m_i/m_e = 4$, $v_{th,e} = 0.5c$, and $v_{th,i} = 0.25c$, was considered. This is the same case as described in Section \ref{sec:sim_perpendicularwaves_double}, with results shown in Figures \ref{fig:kperp_double_Epar_hot}, \ref{fig:kperp_double_Eperp_hot_Ex} and \ref{fig:kperp_double_Eperp_hot_Ey} A total number of 500 particles per cell was used for each species. For strong scaling studies, a mesh size of $1024\times8\times8$ cells was considered, and was kept constant throughout the various runs. For a fixed number of time steps, the simulation was run using 1, 2, 4, 8, 16, 32, 64 and 128 MPI tasks, and the times required for each case were recorded. The metric used to assess strong scaling performance is the speedup, given by
\begin{equation}\label{eq:speedup}
Speedup = t(1)/t(N)
\end{equation}
where $t(1)$ is the amount of time required for the simulation using a single MPI task, and $t(N)$ is the amount of time required for the same simulation using $N$ MPI tasks. The speedup vs number of MPI tasks is shown in Figure \ref{fig:StrongScaling} The speedup stays very close to the ideal speedup up to 32 MPI tasks and starts to deviate from the ideal at 64 MPI tasks and higher. For the weak scaling studies, a mesh size of $64\times8\times8$ cells was considered for a simulation using a single MPI task. For weak scaling, the total problem size per MPI task must be kept constant, as opposed to total absolute problem size as in the case of strong scaling. Hence, for the various subsequent runs performed with 2,4,8,16,32,64 and 128 MPI tasks, the overall problem size was increased in proportion with the number of MPI tasks, with the final run using 128 MPI tasks having a total mesh size of $8192\times8\times8$ cells. The metric used for weak scaling is the efficiency, given by
\begin{equation}\label{eq:efficiency}
\mbox{Efficiency }= t(1)/t(N)
\end{equation}
where $t(1)$ is the amount of time required for the simulation using a single MPI task, and $t(N)$ is the amount of time required for the simulation that is $N$ times larger in terms of problem size when using $N$ MPI tasks. Efficiency vs number of MPI tasks is shown in Figure \ref{fig:StrongScaling}. The high efficiency maintained up to 128 MPI tasks shows good scalability when the problem size per MPI task is kept constant, with only a modest growth in runtime. \nmn{The scalability experiments were performed on a Raven CPU compute node equipped with dual Intel Xeon Platinum 8360Y (Ice Lake-SP) processors, providing 72 physical CPU cores per node, on a distributed-memory architecture with MPI parallelization.} \nmn{Both of these results demonstrate promising scalability and the code's efficient use of the HPC resources within the parameters considered here.}

%%%%%%%%%%%%%%%%%%%%%%%%%%%%%%%%%%%%%%%%%%%%%%%%%%%%%%%%%
\section{Summary, discussion and conclusions}\label{sec:conclusions}
A new numerical scheme based on the geometric, structure-preserving Particle-in-Cell (PIC) model has been developed for the quasineutral, fully kinetic Vlasov-Maxwell equations using mimetic finite differences on dual grids. This capability has been added to the C$++$-based parallel framework, \texttt{GEMPICX}. The scheme has been successfully tested by comparisons between the dispersion relation and results obtained from numerical simulations. This work also demonstrated the capabilities of the scheme in capturing a higher-order process such as damping, through a test case involving the damping of a cyclotron wave. \nmn{Scaling tests have also been conducted, demonstrating encouraging parallel performance on the tested computing platform, although additional studies will be required to assess performance for exascale applications.}

Our model uses a vector potential $\bA$ to obtain the magnetic field using the $\nabla \times \bA$. This results in a curl-curl equation in $\bA$ i.e. equation \eqref{eq:curlcurlA}. We then use the Coulomb gauge condition $\nabla \cdot \bA = 0$ that reduces the curl-curl operator to a Laplace operator resulting in equation \eqref{eq:laplaceA}. This Laplace operator is invertible up to a constant, making it easy to solve for $\bA$ and obtain a unique $\bB$. This, however, is only possible due to the periodic boundary conditions in the test cases considered in this work. The periodic boundaries make the imposition of $\nabla \cdot \bA = 0$ and its discrete counterpart $\tilde{\mathbb{D}} \mathbb{H}_1 \arrA = 0$ straightforward. They also ensure that the discrete de Rham structure is exactly preserved, keeping in line with the idea of structure-preserving methods. However, the use of this module for more realistic scientific applications, such as fusion energy, requires more general, physically accurate non-periodic boundary conditions. With non-periodic conditions, the Coulomb gauge condition $\nabla \cdot \bA = 0$ would not be automatically satisfied and therefore the curl-curl system would not have a unique solution. To obtain meaningful and unique solutions, we would require additional compatibility constraints on $\bA$ and the modification of equation \eqref{eq:laplaceA_disc}. For example, a perfectly conducting wall with unit normal $\bn$ would have $\bA \times \bn = {\bf 0}$, while a perfect magnetic conductor would have $\bn \times ( \nabla \times \bA) = {\bf 0}$. Such non-periodic boundaries would also require appropriate boundary trace spaces within the discrete de Rham structure \citep{arnold2018finite}. The treatment of boundary conditions for the vector potential in a discrete de Rham framework has been extensively discussed in the context of computational electromagnetism \citep{bossavit1998computational, hiptmair2002finite, monk2003finite}, where gauge compatibility and tangential constraints must be handled carefully in non-periodic domains. The principal goal of the current work has been to extend the capabilities of the \texttt{GEMPICX} framework to the quasineutral regime. Introducing non-periodic boundary conditions and the corresponding treatment of the discrete vector potential at boundaries would require a deeper analysis, and this is left for follow-up studies.

\nmn{While the present work only employs second-order Hodge operators, it is worth emphasizing that the mimetic framework naturally admits Hodge discretizations of arbitrary order. Since the Hodge operators determine the spatial accuracy of the discrete Maxwell operator, their order directly influences the accuracy of the underlying field discretization and therefore of wave propagation. In the context of geometric discretizations of the full Vlasov-Maxwell system, higher-order Hodge operators have been shown to improve the representation of short-wavelength modes and reduce numerical dispersion errors, leading to more accurate discrete dispersion relations \citep{kormann2024}. However, the quasineutral approximation considered here removes the high-frequency electromagnetic modes that are most directly affected by such dispersion errors. Consequently, the extent to which higher-order Hodge operators improve the accuracy or efficiency of geometric quasineutral PIC simulations remains an open question. Investigating their impact on low-frequency quasineutral plasma dynamics would require a dedicated study and is left for future work. The numerical implementation of these operators involves denser matrices for the operators themselves and also for linear systems such as equation \eqref{eq:efieldeqn_disc}, increasing computational complexity as compared to the low-order formulation used here.}

Quasineutral fluid models such as magnetohydrodynamics are known to contain sharp gradients and discontinuities such as shocks. Fluid models employ techniques such as limiting, shock-capturing, artificial dissipation and Godunov-type solvers to handle such regimes \citep{toth2000b, balsara1999staggered}. On the other hand, PIC methods, such as the one presented in this work, are not primarily shock-capturing schemes. In regimes where strong gradients or shock-like structures arise, PIC methods represent such features through the underlying particle distribution rather than as true discontinuities in macroscopic fields \citep{birdsall2018plasma}. While the present work focusses on smooth test cases in order to verify the dispersion properties and consistency of the proposed discretization, various PIC studies on shock formation are available in the existing literature \citep{grovselj2024long, birdsall2018plasma, weidl2016hybrid, chen2011energy, Spitkovsky2008}. Geometric discretizations are built using differential and integral operators on the basis of vector calculus identities, and therefore assume smooth fields. Therefore, strong gradients may reduce the accuracy of such spatial discretizations and produce oscillations. In PIC simulations, strong gradients can also amplify particle noise. Numerical stabilization in such regimes typically focuses on improving robustness and controlling particle noise through techniques such as higher-order particle shapes, filtering, or current smoothing \citep{hockney2021computer}. \nmn{We note that structure-preserving approaches for problems with reduced regularity have been investigated in related Eulerian settings, including Hamiltonian and energy-conserving discontinuous Galerkin discretizations for plasma kinetic equations \citep{cheng2014, hakim2020}. However, the treatment of shocks and strongly non-smooth solutions within geometric PIC frameworks remains largely unexplored. Investigating the behavior of the present structure-preserving formulation in strongly nonlinear regimes is therefore an interesting direction for future work.}

While there are several semi-implicit \citep{lapenta2017exactly, markidis2011energy} and fully implicit \citep{chen2015multi, chen2011energy} PIC approaches that enforce energy conservation at the discrete level, our current scheme maintains energy conservation exactly only at the semi-discrete level. After time discretization, energy conservation is only accurate up to the order of the time integrator. This is because the action principle we used for the numerical algorithm was discretized in space but not in time. The time integrator was then obtained by further discretizing in time the semi-discrete governing equations obtained from the action principle. While the focus of this work was extending de Rham structure-preserving methods to quasineutral models, it would be beneficial to extend such methods to exact discrete energy conservation as well. A next logical extension of our method could be to make the method variational in time as well, by discretizing time at the level of the action principle itself, rather than discretizing the resulting equations. Such a fully variational discretization would preserve exact energy conservation after time discretization, significantly enhancing long-term stability \citep{kraus2017gempic, squire2012geometric, marsden2001discrete}.

There are several other interesting avenues for further research. The current work only performs tests that use a Cartesian grid on slab geometry. Application of this model to more realistic magnetic confinement fusion applications such as tokamaks or stellarators, would involve toroidal geometries and therefore require discretizations on curvilinear meshes. This could be achieved by extending the present structure-preserving discretization framework to curvilinear elements, similar to the work of \citet{kreeft2011mimetic}. Another interesting avenue could be the study of wave propagation in a non-constant background magnetic field, which would be of interest to the edge physics community. More recently, \texttt{GEMPICX} has been extended by \citet{meng2025} to the gyrokinetic model in the Zero Larmor Radius limit, namely the drift-kinetic (DK) model \citep{kulsrud1983mhd}. They also coupled the DK model for electrons with the fully kinetic model for ions, resulting in a hybrid drift-kinetic model. A possible future direction for our structure-preserving scheme is its extension to the quasineutral limit of this hybrid drift-kinetic model. Apart from the electromagnetic and Langmuir waves, the quasineutral hybrid model would remove the description of electron cyclotron waves and their harmonics, further reducing time and length scale restrictions on the numerical model.

%%%%%%%%%%%%%%%%%%%%%%%%%%%%%%%%%%%%%%%%%%%%%%%%%%%%%%%%%

\section*{Acknowledgments}
The authors would like to thank Dr. Max Lindqvist for his help with the scaling studies, and Dr. Guo Meng for her fruitful suggestions. Computing resources needed for this work were provided by the Max Planck Computing and Data Facility (MPCDF). This work has been carried out within the framework of the EUROfusion Consortium, funded by the European Union via the Euratom Research and Training Programme (Grant Agreement No 101052200 - EUROfusion). Views and opinions expressed are however those of the author(s) only and do not necessarily reflect those of the European Union or the European Commission. Neither the European Union nor the European Commission can be held responsible for them.

\bibliographystyle{jpp}
\bibliography{gempic-qn}
\end{document}